\documentclass[pra,showpacs,twocolumn]{revtex4}
\usepackage{amsmath,amssymb,epic,eepic}
\usepackage[dvips]{graphicx}
\usepackage{epsfig}
\usepackage{color}
\usepackage{isolatin1}

\begin{document}

\title{Dissipative dynamics of circuit-QED in the mesoscopic regime}

\author{V. Bonzom ${}^{1}$}
\author{H. Bouzidi ${}^{1}$}
\author{P. Degiovanni${}^{1,2}$}
\affiliation{(1) CNRS-Laboratoire de Physique de l'Ecole Normale Sup{\'e}rieure de Lyon, 
46, All{\'e}e d'Italie, 69007 Lyon, France}
\affiliation{(2) Department of Physics, Boston University, 590 Commonwealth avenue, Boston MA 02251, USA}

\begin{abstract}
We investigate the behavior of a circuit QED device when the resonator is initially populated with a mesoscopic 
coherent field. The strong coupling between the cavity and the qubit produces an entangled
state involving mesoscopic quasi-pointer states with respect to cavity dissipation. The overlap of the associated field
components results in collapse and revivals for the Rabi oscillation. Although qubit relaxation and dephasing
do not preserve these states, a simple analytical description of the dissipative dynamics of the circuit QED device
including cavity relaxation as well as qubit dissipation is obtained from the Monte-Carlo approach. 
Explicit predictions for the spontaneous and induced Rabi oscillation signals are derived 
and sucessfully compared with exact calculations. We show that these interesting effects could be 
observed with a 10 photon field in forthcoming circuit QED experiments. 
\end{abstract}


\maketitle 

 \section{Introduction}
 
 Cavity quantum electrodynamics is an ideal playground for the realization of tests of quantum processes,
 simple quantum information processing and exploration of the quantum/classical boundary. 
 Mesoscopic superpositions made of coherent field components with different classical attribues (phase and
 amplitude) and their decay as a result of decoherence
 have been observed in experiments involing circular Rydberg atoms \cite{Brune:1996-2}. 
 Recently, an experimental scheme for cavity QED experiments using
 Josepshon qubits embedded in a superconducting microstrip planar resonator has
 been proposed \cite{Blais:2004-1} and realized
 \cite{Wallraff:2004-1} by the Schoelkopf group at Yale. Relying on conventional lithography
and nanofrabrication techniques, these circuit-QED devices open the way to scalable quantum circuits 
coupling qubits to high-$Q$ cavities acting as quantum buses. Recent experimental 
progresses on these devices have been dramatic, leading to the experimental demonstration 
of quantum non-demolition measurements of the qubit state in the dispersive regime 
\cite{Wallraff:2005-1} and of the resolution of photon number
states of the cavity \cite{Schuster:2007-1}.

\medskip

These recent developements provide a strong motivation toward studying the dynamics of
circuit QED devices at the quantum/classical boundary. In particular, it is important to analyze 
their ability to produce mesoscopic
Schr\H{o}dinger cat states involving coherent components with a mesoscopic number of photons. 
Such states can be produced in the dispersive regime 
where the off resonant qubit acts as a state dependant transparent dielectrics. In this regime,
the interaction between a coherent field and a qubit, initially prepared in a superposition of states naturally
leads to a quantum superposition of phase shifts. 

A much faster way to produce entangled qubit + cavity states is to use 
a resonant qubit + cavity interaction. 
In the mesoscopic regime, photon graininess rapidely 
casts the initially coherent cavity field into a superposition of two components 
with different phases \cite{Gea:1991-1}. This phase splitting is a mesoscopic effect that disappears in 
the classical limit of a very large field, which is then left unaffected by the atoms.
Thus, in the mesoscopic limit, the cavity field acts
as a which-path detector for the atomic states. The overlap of field components of the qubit + cavity state
is responsible for the collapse and revivals of Rabi oscillations which can be either spontaneous or induced by
an echo sequence. 
This resonant phase splitting effect has been evidenced in Rydberg atom experiments for fields containing 
up to a few tens of photons \cite{Auffeves:2003-1}. Its coherence has been checked using an echo 
technique borrowed from NMR \cite{Meunier:2005-1}, following a proposal by 
Morigi {\it et al} \cite{Morigi:2002-1}. 

\medskip

In the presence of dissipation, the entangled qubit + cavity state
is subject to decoherence which leads to a damping of the spontaneous and induced Rabi
oscillation revivals. In a previous work \cite{Degio22}, we 
have developped a simple analytical
model that describes the behaviour of $N$ identical non dissipative qubits resonantly 
and symmetrically coupled to a high-$Q$ cavity. Considering cavity relaxation as the source for decoherence, 
this model is perfectly well suited for describing Rydberg atom experiments. But this is
not sufficient for circuit QED since it does not take into account qubit relaxation and dephasing. 

\medskip

The purpose of the present paper is to present a simple analytical model that takes into account
qubit relaxation and dephasing for a single qubit in a cavity thus generalizing results 
previously obtained by Gea-Banacloche \cite{Banacloche:1993-1} and ourselves \cite{Degio22}. 
Using the physical insight
provided by the stochastic wave function approach \cite{Dalibard:1992-1}, 
an analytic description for the decoherence of the mesoscopic qubit + cavity state is
derived. Our derivation also sheds light on the
range of validity of our previous analysis: whenever the internal interactions within a mesoscopic
device tend to produce pointer states with respect to the dominant coupling to the environment, its
evolution can described in term of these pointer states and decoherence functionals reflecting the 
cumulative imprints they left in the environment. In circuit-QED devices, this simple image
is broken by qubit dissipative processes but a careful analysis of the dynamics enables us 
to obtain simple analytical results for the spontaneous and induced Rabi oscillation signals. 
In principle our analysis can be also used to discuss field tomography and
can also be extended to the case of several qubits.

\medskip

This paper is organized as follows. In Sec. \ref{sec:cQED}, the basic model for cavity-QED is presented,
its dynamics in both resonant and dispersive regimes are briefly recalled. In Sec. \ref{sec:circuit-QED-devices}, 
circuit-QED devices are presented and dissipation sources are discussed. In Sec. \ref{sec:dissipation}, the 
dissipative dynamics of a one qubit circuit-QED device is studied within the stochastic wave function framework. 
For completeness, analytical results are derived both for the resonant and dispersive regimes, recovering in the latter
case results from the Yale group \cite{Gambetta:2006-1}. 
Section \ref{sec:results} presents numerical results obtained from
quantum Monte-Carlo simulations in the resonant regime. These are used to discuss the validity of our 
analytical model and to derive experimentally accessible windows for the observation of mesoscopic qubit + cavity
states in forthcoming circuit-QED devices. In the conclusion, we comment on the inclusion of other
dissipative effects which may be relevant in the study of other cavity QED superconducting 
circuits \cite{Buisson:2001-1}. 

 \section{Cavity QED}
 \label{sec:cQED}

\subsection{The model}
 
 The effective Hamiltonian describing the cavity QED systems is the Jaynes-Cummings
 Hamiltonian \cite{Jaynes:1963-1} involving the coupling of an harmonic mode to
 a two level system (the qubit):
 \begin{equation}
 \label{eq:Jaynes-Cummings}
\hbar^{-1} H_{\mathrm{JC}}=\omega_{0}\,a^\dagger a+ \frac{\omega_{\mathrm{qb}}}{2}\,
 \sigma^z+\frac{g}{2}(a^\dagger\sigma^-+a\,\sigma^+)\,.
 \end{equation}
 In the strong coupling regime, the coupling energy $g$ is assumed to be much larger than
 all energy scales characterizing dissipative processes in the system, {\it i.e.} the cavity relaxation
 rate $\kappa$ and the atomic relaxation and dephasing rates. Neglecting damping, the Jaynes-Cummings 
 Hamiltonian can be exactly diagonalized since the Hilbert space decomposes into two dimensional
 multiplets generated by the states $|e,n\rangle=|+\rangle\otimes |n\rangle$
and $|g,n+1\rangle=|-\rangle\otimes |n\rangle$.  
As a function of the qubit / cavity detuning $\Delta=\omega_{\mathrm{qb}}-\omega_{0}$, the
cavity + qubit system has two different regimes.
 
\subsection{The dispersive regime}

The dispersive regime is reached when the cavity and the qubit are out of
resonance. For an initally empty cavity, it is obtained
for $|\Delta|\gg g$. In this regime, the
eigenstate are very close to the ones of the uncoupled cavity + qubit states.
Performing a second order expansion in $g/\Delta$ leads to 
\begin{equation}
\label{eq:dispersive:Hamiltonian}
\hbar^{-1}H_{\mathrm{disp}}=(\omega_{0}+\chi \sigma^z)\,a^\dagger a+\frac{1}{2}(
\omega_{\mathrm{qb}}+\chi)\,\sigma^z\,.
\end{equation}
where $\chi=g^2/4\Delta$ represent the ac-Stark shift per photon. 
In the presence of a mesoscopic coherent state with average photon number
$\bar{n}$, the cavity + qubit coupling is enhanced
by the coherent field. Off diagonal terms in the Jaynes Cummings multiplet are small 
compared to diagonal ones when $|\Delta|\gg g\sqrt{\bar{n}}$.

\medskip

In the dispersive regime, qubit flips can only occur because of qubit relaxation or external driving of the
system. With respect to the cavity mode, the qubit behaves as a transparent medium with a state-dependent 
refraction index. In Rydberg atom experiments, this regime
has been used to produce Schr\H{o}dinger cat states \cite{Brune:1996-2}, to perform
QND measurement of the photon number \cite{Gleyzes:2007-1} and a measurement of the Wigner
function of the field \cite{Bertet:2002-1}. In circuit QED 
devices, the cavity has been used to performed a QND measurement of the
qubit state \cite{Wallraff:2005-1}. More recently, the strong coupling dispersive
regime where a single photon drastically alters the qubit absorbtion spectrum has been
has been studied theoretically \cite{Gambetta:2006-1} 
and demonstrated experimentally \cite{Schuster:2007-1}. 

\subsection{The resonant regime}

The resonant regime is obtained when $\Delta=0$. In this case, the Jaynes-Cummings eigenstates 
are symmetric and antisymmetric combinations of the form $(|e,n\rangle \pm |g,n+1\rangle)/\sqrt{2}$
separared by an energy $g$ (vacuum Rabi splitting). This leads to vacuum Rabi oscillations
between states $|e,0\rangle$ and $|g,1\rangle$ which 
have been observed \cite{Brune:1996-1} and used to transfer the qubit state in the cavity \cite{Maitre:1997-1}.
In the mesoscopic regime $\bar{n}\gg 1$, 
the resonant interaction leads to an entangled atom-field state with two quasi-coherent field components with 
different classical phases \cite{Gea:1991-1}. The approximate solution by Gea-Banacloche precisely provides
explicit expression for these states and provides the basic framework for discussing the complete dissipative dynamics
of circuit QED devices.

\medskip

It has proven very 
convenient to describe the dynamics using an effective Hamiltonian 
approach that involves effective spin operators acting within the Jaynes-Cummings multiplet
\cite{Klimov:1995-1,Degio22}. Within the framework of this mesoscopic approximation
($\bar{n}\gg 1$ and $t\ll \bar{n}/g$), the evolution of the cQED system can be computed exactly. 
Under the Jaynes-Cummings Hamiltonian, in the mesoscopic regime, the states
\begin{equation}
\label{eq:generalized-GBstates}
|\Psi_{\pm}^X(\theta)\rangle=e^{-\frac{\bar{n}}{2}}\sum_{p=0}^\infty
\frac{\bar{n}^{p/2}}{\sqrt{p!}}\,e^{\mp i\theta\sqrt{p+1}/2}\,|X_{\pm}^{(p)}\rangle
\end{equation}
where $|X^{(p)}_{\pm}\rangle =(|e,p\rangle \pm|g,p+1 \rangle)/\sqrt{2}$
remain of the same form with a time dependent angle $\dot{\theta}(t)=g$. 
In the mesoscopic approximation, $|\Psi_{\pm}^X(\theta)\rangle$ 
can be approximated by a factorized state of the form \cite{Degio22}:
\begin{equation}
\label{eq:GB:factorized-GBstates}
|\Psi_{\pm}^X(\theta)\rangle \simeq e^{\mp i\theta \sqrt{\bar{n}}/2}\,
|D_{\pm}(\theta)\rangle \otimes |\psi_{\pm}(\theta)\rangle
\end{equation}
where the atomic polarization is given by:
\begin{equation}
\label{eq:GB:atomic-polarizations}
|D_{\pm}(\theta)\rangle=\frac{1}{\sqrt{2}}
\left(\pm e^{\mp i\theta}|+\rangle + |-\rangle\right)
\end{equation}
and the field component is
\begin{equation}
\label{eq:GB:field-states}
|\psi_{\pm}(\theta)\rangle= e^{\pm i\theta\sqrt{\bar{n}}/2}\,
e^{-\bar{n}/2}\sum_{k=0}^{+\infty}
\frac{\bar{n}^{k/2}}{\sqrt{k!}}\,e^{\mp i\theta \sqrt{k}/2}\,|k\rangle\,.
\end{equation} 
The field state $|\psi_{\pm}(\theta)\rangle $
is proportional to a quasi coherent state whose parameter is roughly equal to
$\sqrt{\bar{n}}\,e^{\mp i\theta/4\sqrt{\bar{n}}}$. At fixed $\bar{n}$, 
the phase, in Fresnel plane, associated with the generalized 
Gea-Banacloche state 
$|\Psi^X_m(\theta)\rangle$ is $\phi_{m,\theta}=-m \theta/2\sqrt{\bar{n}}$.
The atomic
polarization evolves in the equatorial plane of the Bloch spere, its phase being perfectly
correlated to the electromagnetic phase. 

\medskip

The states $|\psi_{\pm}(\theta)\rangle$ act as a 
path detector for the atomic polarizations $|D_{\pm}(\theta)\rangle$ and this explains the pattern
of collapses and revivals of Rabi oscillations. When $t\ll g^{-1}$, the field components 
$|\psi_{\pm}(gt)\rangle$ still overlap and therefore, Rabi oscillations arising from interference
between the $|D_{+}\rangle$ and $|D_{-}\rangle$ are observed. Once $gt\gtrsim 2\pi$, the 
field components do not overlap and Rabi oscillations disappear. Only when
$gt/2\sqrt{\bar{n}}$ is a multiple of $2\pi$, the two states overlap again. The information about 
the qubit state stored by into the cavity is erased, thus leading to a revival of Rabi 
oscillations. Henceforth, collapses and revivals of Rabi oscillations in the mesoscopic
regime are a direct illustration of the complementarity principle.

When $gt/2\sqrt{\bar{n}}=\pi$, the  qubit polarizations will coincide. Then, 
the atom + cavity state disentangles and the resulting 
cavity state is a mesoscopic Schr\H{o}dinger cat states involving two quasi-coherent
components with opposite phases. This occurs much faster than in the
dispersive regime $|\Delta|\gg g\sqrt{\bar{n}}$ where such a Schr\H{o}dinger 
cat state would be generated in $\pi|\Delta|/g^2\gg 2\pi\sqrt{\bar{n}}/g$. 

 \section{Circuit QED devices}
 \label{sec:circuit-QED-devices}

Recent experiments performed at Yale involve the coupling of a superconducting Josephson 
qubit and a superconducting microstrip resonator with high quality factor
\cite{Wallraff:2004-1}. We will first present these devices and recall how they
can be described by the Jaynes-Cummings Hamiltonian. Dissipation sources will
then be discussed.

\subsection{The cavity: a superconducting microstrip resonator}

The resonator is 1D a transmission line made by photolithography whose lowest 
mode lies within the 1-10~GHz frequency range and whose quality factors can be 
as high as $10^6$ \cite{Frunzio:2005-1}. 
Besides this, the electromagnetic field 
associated with their eigenmodes is confined within a relatively small volume, thus leading 
to high electric fields between the center and ground planes (typically $0.2$~V/m). 
Such a resonator is usually characterized by a wave velocity $v$, a real impedance $R$ ans its length $L$.

The finite length superconducting resonator has many stationary modes up to a certain
high frequency cutoff but in the present situation, only the coupling to the lowest energy mode
The voltage difference between the inner and outer electrodes at point $x$ is then expressed in
terms of the mode creation and destruction operators:
\begin{equation}
\label{eq:yale:V}
V(x)= V_{0}\,(a+a^\dagger)\,\sin{\left(\frac{\pi x}{L}\right)}
\end{equation}
where $V_{0}$ denotes the maximum voltage felt by the qubit.

\subsection{The artificial atom: a Josephson qubit}

Within the resonator lies a superconducting Josephson charge qubit (initially a Cooper pair box)
coupled to the cavity electric field. It consists in a small superconducting island connected 
to a superconducting reservoir through a very thin insulating barrier.
When the temperature $k_{B}T$ and the
the Coulomb charging energy of the island $E_{c}$ are smaller than the superconducting gap, 
the qubit degrees of freedom are encoded by the charge 
state of the superconducting island. 
This device can be characterized by the Josephson amplitude $E_{J}$ and the Coulomb energy 
$E_c$. Control parameters are the voltage potential
$V_{g}$ imposed to the superconducting island and the Josephson amplitude which can be tuned 
using a SQUID instead of a single Josephson junction \cite{Makhlin:2001-1}: 
$E_{J}=\mathcal{E}_{J}\,\cos{(2\pi\Phi_{J}/\Phi_{0})}$ where $\Phi_{0}=2e/h$ denotes
the flux quantum and $\Phi_{J}$ the magnetic flux through the SQUID.

In the charge regime $E_{c}\ll E_{J}$ and for gate charge $n_{g}=C_{g}V_{g}/2e$
restricted to a unit charge interval $0\leq n_{g}\leq 1$, the system effectively behaves like a two
level system (TLS) involving two adjacent charge states of the island. In this charge basis, its effective 
Hamiltonian is given by:
\begin{equation}
\label{eq:yale:Hqb}
H_{\mathrm{qb}}=-\frac{B_{z}}{2}\,\sigma^x-\frac{B_{x}}{2}\,\sigma^z
\end{equation}
where $B_{z}=4E_{c}(1-2n_{g})$ and $B_{x}=E_{J}$.

\subsection{The qubit/cavity coupling}

The electromagnetic mode of the 1D cavity naturally couples to the qubit charge through capacitance
between the resonator inner and outer electrodes. The coupling $g$ energy is limited 
by the dipolar energy of the qubit within the resonator:
$g/\omega_{0}\lesssim \lambda_{\mathrm{c}}\sqrt{R/R_q}$ where $\lambda_{\mathrm{c}}$ is a capacitance
ratio and $R_q=h/e^2$ is the quantum of resistance. Classical electrodynamics shows that $R$ is of the order of
$2R_q\alpha\Lambda/\epsilon_{r}$ where $\alpha$ is the fine structure constant,
$\epsilon_{r}$ is the relative permittivity and
$\Lambda$ a geometric factor, usually logarithmic in the microstrip
aspect ratios. Within the 1 to 10 GHz frequency range, values of 
$g/2\pi$ range from 11 to 105~MHz  (see table \ref{table:1}) the latter value being obtained using
the "transmon" \cite{Koch:2007-1}. 
Assuming that the two level description of the qubit is valid, the coupling Hamiltonian between 
the qubit and the cavity is given by:
\begin{equation}
\label{eq:Yale:Hint}
H_{\mathrm{int}}=\hbar g\,(a+a^\dagger)(1-2n_{g}-\sigma^x)\,.
\end{equation}
The coupling being small compared to the resonator and qubit eigenfrequencies,
the rotating wave approximation is valid and
close to the charge degeneracy point $n_{g}\simeq 1/2$, eq. \eqref{eq:Yale:Hint} 
reduces to the Jaynes-Cummings Hamiltonian \eqref{eq:Jaynes-Cummings}.
 
 \subsection{Measurement protocols}
 
In these experiments, the Josephson qubit cannot be measured directly. This circuit-QED
system is probed through the resonator which is then connected to external ports
at its ends. 
A network analyzer is used to analyzed the phase and
amplitude of a classical electromagnetic wave transmitted through the device. This method
has been used to perform a non-destrictive measurement of the qubit state using a dispersive measurement 
technique \cite{Wallraff:2005-1}. Thus, the relaxation time scale of the resonator $\kappa^{-1}$  
sets a lower bound on the measurement time. 
Probing the state of the qubit thus requires that $\kappa \gtrsim \gamma_{2}$ where $\gamma_{2}=
\gamma_{1}+\gamma_{\varphi}/2$. Although cavities with quality factors of the order of $10^6$ have 
been manufactured \cite{Frunzio:2005-1}, 
devices used in experiments have a lower $Q$ in order to satisfy the fast measurement constraint. 

\medskip

In order to avoid this limitation, one has to rely on an
alternative method for measuring the qubit state. Recently, a new detection scheme 
based on the dynamical bifurcation of Josephson junction has been realized and
provides a high contrast, low backaction, dissipationless and fast measurement of the qubit state
\cite{Siddiqi:2005-1}. Thus, rapid improvement of Josepshon circuit technology suggests that alternative
methods of detection might become available in the near future.

In the present paper, we focus on the evolution of the qubit + cavity system prior
to measurement. In particular, our main objective is to describe the dynamics of the circuit QED
system at resonance in the mesoscopic regime and shed light on the main physical
effects independantly from the measurement protocol. Therefore, for the sake of simplicity, 
we will not attempt modeling the subsequent measurement process. 
 
 \subsection{Dissipation mechanisms}
 
 \subsubsection{Qubit relaxation and dephasing}
 
Qubit dissipation and decoherence arise from their coupling to environmental degrees of freedom
either extrinsinc (measurement circuit) or intrinsic (structural defects inside the material). Relaxation
involves energy exchange between the qubit and its environment and, as such, is generically sensitive to
the low frequency part of the environmental spectrum. Relaxation also leads to decoherence defined as the 
decay of off diagonal matrix element in the qubit's eigenbasis. 

But dephasing can also occur without
energy exchange (pure dephasing). Early circuit-QED experiments used a Cooper pair box. 
The coherence properties
of these devices are limited by pure dephasing induced by low frequency noise. It arises from fluctuating charges
in the insulating amorphous $\mathrm{Al}_{2}\mathrm{O}_{3}$ used to make the Josephson 
 junctions \cite{Astafiev:2005-1}. It is known that the effect of this low frequency 
 noise cannot be described within
a Markovian framework \cite{Makhlin:2003-1}. A possible escape to this problem is to 
operate the qubit at special point where sensitivity to voltage fluctuations is at second order instead of first
order \cite{Vion:2002-1}. At this working point, treating the effect of low frequency noise 
requires going beyond the Markovian approximation \cite{Makhlin:2004-2,Ithier:2005-1}.
Fortunately recent circuit QED experiments use another type of qubit, called the "transmon" which
has very low sensitivity to voltage fluctuations (sweet spot everywhere), 
can be operated as a two level system and the low frequency noise 
\cite{Koch:2007-1}. Therefore, having in mind future experiments 
performed with transmons,  qubit relaxation 
and dephasing will be treated within the Markovian approximation in the present paper.

\medskip

Under this hypothesis, the evolution of the qubit + cavity system reduced density operator is
described by a master equation of the form:
\begin{equation}
\label{eq:master-equation}
\frac{d\rho}{dt} = -\frac{i}{\hbar}\,[H,\rho] + \sum\mathcal{L}_{j}(\rho)
\end{equation}
where $H$ denotes the Jaynes-Cummings Hamiltonian \eqref{eq:Jaynes-Cummings} and
$\mathcal{L}_{j}(\rho)$ are the Lindbladian superoperators associated with each markovian
dissipation channel. 

\medskip 

The relaxation Lindbladian operator is given by:
 \begin{equation}
\mathcal{L}_{r}(\rho)=\gamma_{1}\,\left(\sigma^-\ldotp \rho \ldotp \sigma^+ 
-\frac{1}{2}\,\left\{\frac{1+\sigma^z}{2},\rho\right\}\right)\,.
\end{equation}
where $\gamma_{1}$ represents the relaxation rate of the isolated qubit. 
Markovian pure dephasing corresponds to the difusion of the relative phase between states $|\pm\rangle$. 
The pure dephasing Lindbladian operator is then given by:
\begin{equation}
 \mathcal{L}_{r}(\rho)=\frac{\gamma_{\varphi}}{2}\left(\sigma^z\ldotp \rho\ldotp \sigma^z-\rho\right)\,.
 \end{equation}
 where $\gamma_{\varphi}$ is the pure dephasing rate of the qubit defined as the decaying rate
 of the $\langle +|\rho(t)|-\rangle$ off diagonal matrix element for a standalone qubit. 
 
\subsubsection{Resonator relaxation}

In the case of the circuit-QED devices, relaxation mainly comes from
capacitives losses at the ends of the resonator. The temperature dependance can be
explained by considering that the inverse quality factor $Q^{-1}$ is a sum of two contributions.
The first contribution represents thermal breaking of Cooper-pairs in the superconductor and it 
scales as $\exp{(T_{c}/T)}$. The other contribution
has a much weaker temperature dependance and is probably associated with intrinsic dissipation
mechanisms such as dielectric losses or magnetic vortices \cite{Frunzio:2005-1}. 
This term is responsible for the  low temperature value of
quality factor.
Note that the current limitation in the quality factor of microwave cavities used in 
Rydberg atom experiments comes from diffraction losses, surface rugosity and residual
resistance of the superconducting mirrors coming from defects, impurities and
magnetic vortices \cite{Kuhr:2006-1}.

The corresponding Lindbladian is then:
\begin{equation}
\mathcal{L}_{c}(\rho)=\kappa\,\left(a\ldotp \rho \ldotp a^\dagger -\frac{1}{2}(
a^\dagger a\ldotp\rho+\rho\ldotp a^\dagger a)\right)\,.
\end{equation}
where $\kappa$ denotes the total relaxation rate of the oscillator.

 \section{Dissipative dynamics}
 \label{sec:dissipation}

\subsection{The stochastic wave function method}

In principle, eq. \eqref{eq:master-equation} can be solved
numerically in order to obtain the quantum dynamics. 
However, an analytical ansatz for the reduced density matrix can be
found within the mesoscopic approximation using the 
the quantum jump approach \cite{Dalibard:1992-1} to the dissipative
dynamics of the atoms + cavity system. As we shall see extensively in this paper, it
provides a deep and useful insight into the full dissipative dynamics of the cQED system
and will enable use to find simple analytical results for the Rabi oscillation signals.

\medskip

The quantum jump method provides a solution to the master equation \eqref{eq:master-equation} 
by assuming
that the environment of the system is
continuously monitored so that any emission or absorption of quanta by the system
can be assigned a precise, although stochastic date. 
Each time such an event occurs, the system undergoes a quantum
jump: 
\begin{equation} 
|\psi(t^+)\rangle = \frac{L_j\,|\psi(t)\rangle}{\sqrt{
\langle L_j^\dagger L_j\rangle_{|\psi(t^-)\rangle}
}}\,.
\end{equation}
The probability rates at a given time $t$ for the various quantum 
jumps are directly obtained as averages 
$\langle L_{j}^\dagger L_{j}\rangle
_{|\psi(t^-)\rangle}$
where the $L_j$ denote the quantum jump operator of type $j$. 
Here, the quantuum jump
operators associated to pure dephasing, qubit relaxation and cavity relaxation are
respectively equal to: $L_{\varphi}=\sqrt{\gamma_{\varphi}/2}\,
i\sigma^z$ corresponding to a $\pi$ rotation around the $z$ axis, 
$L_{R}=\sqrt{\gamma_{1}}\,\sigma^-$ and $L_{\mathrm{c}}=\sqrt{\kappa}\,a$.
 
Between these jumps, the evolution is described by an effective Hamiltonian that
describes both its intrinsic dynamics and the acquisition of information arising from
the fact that no quanta has been detected:
\begin{equation}
\label{eq:nh-hamiltonian}
H_{\mathrm{eff}}=H-\frac{i\hbar}{2} \sum_{j}\,
L^\dagger_jL_j\,.
\end{equation}
The reduced density matrix
is then recovered by averaging over the set of stochastic trajectories associated 
with a large set of quantum jumps sequences. The weight
of a given trajectory can be directly related to the dates and 
types of the various quantum jumps. The corresponding Linbladian is
then given by:
\begin{equation}
\label{eq:lindbladian-L}
\mathcal{L}_{j}(\rho)=L_{j}\ldotp \rho\ldotp L_{j}^\dagger -
\frac{1}{2}\lbrace L_{j}^\dagger L_{j},\rho\rbrace\,.
\end{equation}
This method proves to be very
convenient numerically since the number of variables involved is of the order
of the dimension $d$ of the system's Hilbert space whereas it scales
as $d^2$ in the master equation approach. 

\subsection{Pointer states dynamics}
 \label{sec:dissipation:pointer-states}

A general stochastic dynamics tends to produce arbitrary mixtures of states. However,
it takes a very simple form for states that are preserved (up to a phase) 
between and during quantum jumps. These
so called pointer states entangle minimally with the system's environment. In some cases,
they provide a basis of the system's Hilbert space, thus leading to a simple description
of the stochastic dynamics which we call a pointer state dynamics. 

Since coherent states are pointer
states with respect to cavity losses, such a description
is directly relevant for the describing the effect of cavity relaxation on cavity QED systems 
in both dispersive and resonant regime.

\subsubsection{General results}
 \label{sec:dissipation:pointer-states:general}

We assume that each eigenvalue of the quantum jump operator $L$ is 
non degenerate and choose a fixed section
$\lambda \mapsto |\lambda\rangle$ ($L|\lambda\rangle=\lambda\,|\lambda\rangle$)
of the corresponding vector bundle over $L$'s spectrum. 
Pointer states $|\lambda\rangle$ remain pointer states in the
evolution between quantum jumps as soon as
$[L,H]=h(L)$ and $[L,L^\dagger L]=f(L)$ (sufficient condition). 
In this case, evolution beween quantum jumps
of $|\lambda\rangle $ produces a single quantum trajectory $t\mapsto e^{i\theta(\lambda,t)}\,
|\lambda(t)\rangle$ where:
\begin{equation}
\label{eq:pointer:equadiff}
\frac{d\lambda(t)}{dt}=-\frac{i}{\hbar}\,h(\lambda(t))-\frac{\gamma}{2}\,f(\lambda(t))
\end{equation}
with initial condition $\lambda(0)=\lambda$.
The phase $\theta(\lambda,t)$ contains an Hamiltonian and a Berry phase contribution:
\begin{eqnarray}
\label{eq:pointer:phase}
\theta_{t}[\lambda] & = & 
-\frac{1}{\hbar}\int_{0}^t\langle \lambda(\tau)|H|\lambda(\tau)\rangle\,d\tau \\
 & + & \int_{0}^t\Im{(\langle\dot{\lambda_{\tau}}|\lambda(\tau)\rangle)}\,d\tau
\end{eqnarray}
Then, starting from a single pointer state $|\lambda\rangle$, the stochastic dynamics
taking into account all possible sequences of quantum jumps occuring at times
$0\leq t_{1}\leq \ldots \leq t_{p}\leq t$ produces a single
trajectory $t\mapsto \lambda(t)$ but encodes the sequence of quantum jumps in a 
$(t_{1},\ldots ,t_{p})$ dependent phase. The resulting state at time $t$ is of the form
$e^{i\theta(t_{1},\ldots,t_{p})}|\lambda(t)\rangle$ and
therefore, any initial state which is linear combination of pointer states will 
experience decoherence because of the averaging of these random phases. 

Summing over all quantum jumps sequences leads to 
the system's reduced density operator obtained from
an initial state $\sum_{\lambda}c_\lambda\,|\lambda\rangle$:
\begin{equation}
\label{eq:pointer:rho}
\rho(t)=\sum_{(\lambda_{+},\lambda_{-})}
c_{\lambda_{+}}\,c_{\lambda_{-}}^*\,\mathcal{F}_{t}[\lambda_{+},\lambda_{-}]
e^{i(\theta_{t}[\lambda_{+}]-\theta_{t}[\lambda_{-}])}
|\lambda_{+}(t)\rangle\langle\lambda_{-}(t)|
\end{equation}
where the decoherence functional is given by:
\begin{equation}
\label{eq:pointer:decoherence}
\mathcal{F}_{t}[\lambda_{+},\lambda_{-}]=
e^{-\frac{\gamma}{2}\int_{0}^t|\lambda_{+}-\lambda_{-}|^2(\tau)\,d\tau}
e^{i\gamma \int_{0}^t\Im{(\lambda_{+}\lambda_-^*)(\tau))}\,d\tau}\,.
\end{equation}
As expected, this expression coincides with the accumulated decoherence of a pair
of coherent states driven along trajectories $t\mapsto \lambda_{\pm}(t)$. 

\subsubsection{Application to cQED systems}
 \label{sec:dissipation:pointer-states:cQED}

In the dispersive regime of cQED,
the effective dissipative dynamics is described by the effective Hamiltonian 
\eqref{eq:dispersive:Hamiltonian}. In the strong coupling regime $g\ll \kappa$, the 
effective quantum jump operator associated with 
cavity losses can still be approximated by $L=\sqrt{\kappa}\,a$. Therefore, in this scheme, 
the pointer states with respect to cavity losses are states of the form
$|\pm\rangle\otimes|\alpha\rangle$. Since any state can be expanded as a linear combination
of these pointer states, the dispersive regime of cQED realizes the above describes pointer
state dynamics. Their respective coherent state parameters evolve according to:
\begin{equation}
\label{eq:pointer:cQED:dispersive:eqlambda}
\frac{d\alpha_{\pm}}{dt}=-i\omega_{\pm}\,\alpha_{\pm}(t)-
\frac{\kappa}{2}\,\alpha_{\pm}(t) -i\epsilon(t)
\end{equation}
where $\omega_{\pm}=\omega_{0}\pm\chi$ is the ac-Stark shifted cavity frequency
and $\epsilon(t)$ is a driving of the cavity. Starting from
an initial state of the form $|\psi_{\mathrm{qb}}\rangle \otimes |\alpha\rangle$ leads
to a qubit + cavity reduced density operator of the form \eqref{eq:pointer:rho} involving
two pointer states $|+\rangle \otimes |\alpha_{+}(t)\rangle$ and 
$|-\rangle \otimes |\alpha_{-}(t)\rangle$ where
$\alpha_{\pm}(\tau)$ denote solutions of \eqref{eq:pointer:cQED:dispersive:eqlambda} 
with initial condition $\alpha_{\pm}(0)=\alpha$.
The decoherence functional induced by cavity losses and relative to these
states is obtained by substituting $\alpha_{\pm}(\tau)$
into \eqref{eq:pointer:decoherence}. As we shall see in more details in 
section \ref{sec:dissipation:relaxation:dispersive}, 
this is exactly the dynamics solved in Ref. \cite{Gambetta:2006-1} using a different method.

\medskip

In the case of a qubit resonantly coupled to the cavity, generalized Gea-Banacloche states
\eqref{eq:generalized-GBstates}
are approximate pointer states with respect to cavity losses. Let us first consider 
an initial state of the form $|\psi_{\mathrm{qb}}\rangle\otimes |\alpha\rangle$ where $|\psi_{\mathrm{qb}}\rangle$
is the qubit initial state and where the parameter $\alpha $ of the initial coherent field
is taken real.
Cavity losses will then produce a qubit + cavity reduced density operator involving two generalized
Gea-Banacloche states $|\Psi_{\pm }^X(t)\rangle$. The decoherence coefficient
in front of the $|\Psi_{+}^X(t)\rangle \langle \Psi^X_{-}(t)|$ is obtained by 
substituting $\alpha_{\pm}(t)=\alpha \,e^{\mp igt/4\sqrt{\bar{n}}}$ in
\eqref{eq:pointer:decoherence}. Introducing one or several $\pi$-pulses (single echo
or bang-bang control) only changes the trajectories of the Gea-Banacloche parameters
$\alpha_{\pm}(t)$ in the Fresnel plane but eq. \eqref{eq:pointer:decoherence} remains valid.
To deal with more general initial condition such as $|\psi_{\mathrm{qb}}\rangle\otimes
(|\alpha_{1}\rangle+|\alpha_{2}\rangle)$,
one just has to decompose the initial state as a sum of rotated generalized Gea-Banacloche states
where the $X$ direction is replaced by the directions given by the phases of the coherent states
present in the initial condition. 

\medskip

So far, only the effect of cavity losses has been discussed. How do other sources of dissipation affect
this picture in the dispersive and the resonant regimes~? 

\medskip

To answer this question, let us first remark that, in the dispersive regime,
pointer states with respect to cavity losses are also pointer states with respect to pure dephasing of the qubit.
Qubit relaxation sends a pointer state of  the form $|+\rangle\otimes |\alpha\rangle$
on a pointer state $|-\rangle\otimes |\alpha\rangle$. Therefore, in the dispersive regime, a pointer state
is sent on another pointer state. But the situation is more involved in the resonant or quasi-resonant regimes.
However, we shall see that quasi pointer states are sent on linear combinations of quasi pointer states.
Thus, the complete stochastic dynamics can still be described in terms of generalized Gea-Banacloche states
\eqref{eq:generalized-GBstates} and for this reason, simple analytical results can
still be obtained. 

The simpler case of pure dephasing will be considered first and we will then turn
to the more complicated case of qubit relaxation. Explicit analytical results for the spontaneous and
induced Rabi oscillation signals are given in section \ref{sec:dynamics:Rabi-results}.
For the sake of simplicity, the stochastic dynamics
will be discussed using the factorized form \eqref{eq:GB:factorized-GBstates} 
of generalized Gea-Banacloche states, assuming that
their coherence is not broken by the dissipative processes under consideration. This
assertion will be justified by a more precise discussion of the dynamics postponed in 
appendix \ref{appendix:dynamics}.
   
\subsection{Pure dephasing}
 \label{sec:dissipation:dephasing}
  
\subsubsection{Resonant regime}
 \label{sec:dissipation:dephasing:resonant}

First of all, let us note that
the corresponding quantum jump operator has no effect between quantum jumps.
Moreover, the rate of pure dephasing jumps is constant in time, independent of the
state and equal to $\gamma_{\varphi}/2$. 
Next, each quantum jump acts as a $\pi$ rotation around 
the $z$ axis, exactly like an echo pulse. As such, it sends the atomic polarization
$|D_{\pm}(\theta)\rangle$ defined in eq. \eqref{eq:GB:atomic-polarizations}
on $|D_{\mp}(-\theta)\rangle$ not affecting the field component
\eqref{eq:GB:field-states}. After $p$ jumps occuring at times 
$0\leq t_{1}\leq \ldots \leq t_{p}\leq t$,
an initial state $|D_{m}(0)\rangle \otimes |\psi_{m}(0)\rangle$ has turned
into $e^{-im\theta\sqrt{\bar{n}}}\,
|D_{m'}(\theta_{t})\rangle \otimes |\psi_{m'}(\theta_{t})\rangle$
where $m'=(-1)^pm$ and $\theta_{t}=\theta(t_{1},\ldots ,t_{p};t)$ is 
defined as
\begin{equation}
\theta(t_{1},\ldots ,t_{p};t)=g\sum_{j=0}^p(-1)^j(t_{j+1}-t_{j})
\end{equation}
where $t_{0}=0$ and $t_{p+1}=t$. 

\medskip

\begin{figure}
\begin{center}
\begin{picture}(0,0)%
\epsfig{file=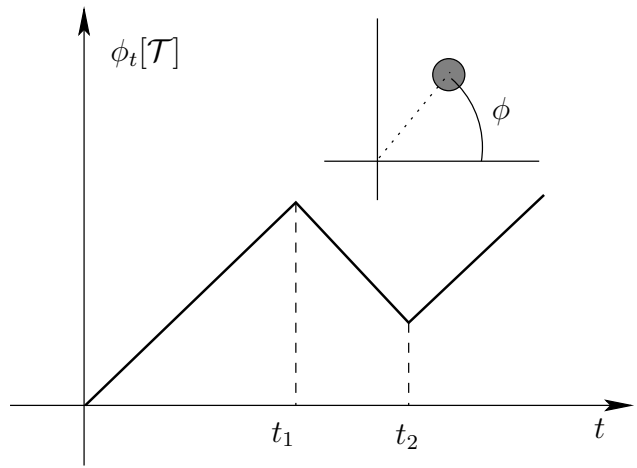}%
\end{picture}%
\setlength{\unitlength}{4144sp}%
\begingroup\makeatletter\ifx\SetFigFont\undefined%
\gdef\SetFigFont#1#2#3#4#5{%
  \reset@font\fontsize{#1}{#2pt}%
  \fontfamily{#3}\fontseries{#4}\fontshape{#5}%
  \selectfont}%
\fi\endgroup%
\begin{picture}(3759,2776)(3139,-3583)
\put(3751,-1148){\makebox(0,0)[lb]{\smash{{\SetFigFont{12}{14.4}{\familydefault}{\mddefault}{\updefault}{\color[rgb]{0,0,0}$\phi_t[\mathcal{T}]$}%
}}}}
\put(6643,-3407){\makebox(0,0)[lb]{\smash{{\SetFigFont{12}{14.4}{\familydefault}{\mddefault}{\updefault}{\color[rgb]{0,0,0}$t$}%
}}}}
\put(4713,-3427){\makebox(0,0)[lb]{\smash{{\SetFigFont{12}{14.4}{\familydefault}{\mddefault}{\updefault}{\color[rgb]{0,0,0}$t_1$}%
}}}}
\put(5457,-3441){\makebox(0,0)[lb]{\smash{{\SetFigFont{12}{14.4}{\familydefault}{\mddefault}{\updefault}{\color[rgb]{0,0,0}$t_2$}%
}}}}
\put(6032,-1497){\makebox(0,0)[lb]{\smash{{\SetFigFont{12}{14.4}{\familydefault}{\mddefault}{\updefault}{\color[rgb]{0,0,0}$\phi$}%
}}}}
\end{picture}%
\end{center}
\caption{\label{fig:dephasing-trajectory}
Typical trajectory of the phase $\phi$ of the quasi-coherent component state in the Fresnel plane
showing two quantum jumps at time $t_{1}$ and $t_{2}$. The effect of a pure dephasing quantum
jump is to reverse the direction of motion of the quasi-coherent component in the Fresnel
plane. Inset recalls the definition of $\phi$.}
\end{figure}

A typical trajectory $\mathcal{T}$ of the associated phase in Fresnel plane 
is depicted on figure \ref{fig:dephasing-trajectory}, with slope changing of
sign at each quantum jump. 
The relative phase between the two quasi-coherent field components follows a random
trajectory directly related to the integral of a telegraphic noise:
\begin{equation}
(\Delta\phi)(t)=\frac{g}{2\sqrt{\bar{n}}}\,\int_{0}^tX(\tau)\,d\tau
\end{equation}
where $X(\tau)\in\{-1,1\}$ starts with $X(0)=1$ 
and jumps each time there is a quantum jump.
The probability distribution of waiting times $\tau$ between quantum jumps
is exponential $\psi(\tau)=\frac{\gamma_{\varphi}}{2}\,e^{-\gamma_{\varphi} \tau/2}$.
The characteristic function of the probability distribution for $(\Delta\phi)(t)$ is
nothing but the decoherence coefficient of a qubit in a 
longitudinal telegraphic noise \cite{Paladino:2002-1}.
Its behavior depends on the dimensionless parameter 
$\eta=g/\gamma_{\varphi}\sqrt{\bar{n}}$ which is equal to 
$2\pi/\overline{N}_{\varphi}$ where
$\overline{N}_{\varphi}=2\pi\sqrt{\bar{n}}\gamma_{\varphi}/g$ 
is the average number of quantum jump events occuring before the date 
of the first expected Rabi oscillation revival.

\medskip

For $\eta \gg 1$, only a single event is necessary to spread the phase 
 $(\Delta\phi)(t)$ over $2\pi$. In this regime, the evolution of the cQED system
 up to the first revival 
 is dominated by quantum trajectories having no pure dephasing quantum 
 jumps. Their total weight is $e^{-\gamma_{\varphi} t/2}\sim 1$. Trajectories
 with one quantum jump or more have a total weight $1-e^{-\gamma_{\varphi} t/2}\ll 1$ 
 and can therefore be treated perturbatively.
 For fixed $g$ and $\gamma_{\varphi}$, 
 this "weak dephasing limit" is realized for $\bar{n}\ll (g/\gamma_{\varphi})^2$. In practice, this
 is the regime of interest for present and forthcoming experiments since $g/\gamma_{\varphi}$ 
 is above 40. 
 
\medskip
 
For $\eta \ll 1$, a large number of events are necessary to spread significantly the
random relative phase $(\Delta\phi)(t)$. 
In this case, rapid dephasing of the qubit prevents the observation of an atom + cavity
entangled state. As shown in appendix 
\ref{appendix:dynamics}, the incoherent qubit ends up breaking the
coherence of the coherent field, selecting Fock states as pointer states.
This is the "strong dephasing limit" which, as already stressed, can only be reached for very
large photon numbers in the strong coupling regime of cQED. 

Nevertheless, in the strong dephasing limit, even if the spreading of the field phase takes place
over a time scale $2\gamma_{\varphi}\bar{n}/g^2\gg \gamma_{\varphi}^{-1}$
the Rabi oscillation signal will be strongly damped in a much shorter
time. Indeed for $t\lesssim 2\pi\sqrt{\bar{n}}/g$, 
although the Fresnel angle of Gea-Banacloche states are weakly dispersed around zero
and thus have a rather strong overlap for a typical given quantum trajectory, 
the classical Rabi oscillation phase $e^{2i\bar{n}\phi_{t}}$
averages to zero
over a time scale of the order of $2\gamma_{\varphi}^{-1}$. This means that
the qubit dephasing time is anyway the upper limit to the Rabi oscillation visibility in
the mesoscopic regime. Nevertheless, for qubits in high quality resonators but rather poor
dephasing properties $g/\gamma_{\varphi}\lesssim 5$, the crossover from weak to strong
dephasing regimes might be observed by performing a tomography of the field state.

\medskip

Finally, the result of this analysis is that, in the strong coupling regime of cQED, 
the main contribution to the Rabi oscillation signal comes from quantum histories without
any pure dephasing quantum jump. As far as we are interested by this signal, the effect of qubit
dephasing can be accounted for through a supplementary decoherence factor on top of the
decoherence factor associated with cavity losses:
\begin{equation}
\mathcal{F}_{+,-}^{(\varphi)}(t)\simeq e^{-\gamma_{\varphi}t/2}\,.
\end{equation}

 \subsubsection{Dispersive regime}
  \label{sec:dissipation:dephasing:dispersive}

In the dispersive regime, the analysis is much simpler since pointer states with respect
to cavity losses are also pointer states with respect to pure dephasing of the qubit. Therefore,
starting with an initial state of the form $|\psi_{\mathrm{qb}}\rangle \otimes
|\alpha\rangle$, the only effect of pure dephasing is to multiply the decoherence functional 
associated with cavity losses by:
\begin{equation}
\mathcal{F}_{+-}^{(\varphi)}(t)= e^{-\gamma_{\varphi} t}\,.
\end{equation}

\subsection{Qubit relaxation}
 \label{sec:dissipation:relaxation}

\subsubsection{Resonant regime}
 \label{sec:dissipation:relaxation:resonant}

For a standalone qubit, 
the non hermitian part of the Hamiltonian tends to bring the state of the qubit towards 
the $|-\rangle$ state, thus reflecting the acquisition of information associated with the
absence of relaxation quantum jump. But at resonance, in the presence of a mesoscopic state, the pumping 
of the qubit by the cavity photons alters the dynamics between quantum jumps.
As discussed in appendix \ref{appendix:non-hermitian}, 
in the $\gamma_{1}\ll g$ regime, the dynamics is dominated by the 
strong atoms + cavity coupling. Contrarily to the case of a the dispersive
regime where no photon can excite the qubit after relaxation, 
several relaxation jump can occur. The statistics of
relaxation quantum jumps is indeed described by a renewal process
with effective waiting time distribution $\psi_{1}(\tau)=(\gamma_{1}/2)e^{-\gamma_{1}\tau/2}$.
The effective rate $\gamma_{1}/2$ follows from
the localization of generalized Gea-Banacloche states in the 
equatorial plane of the Bloch sphere.

\medskip

\begin{figure}
\begin{picture}(0,0)%
\epsfig{file=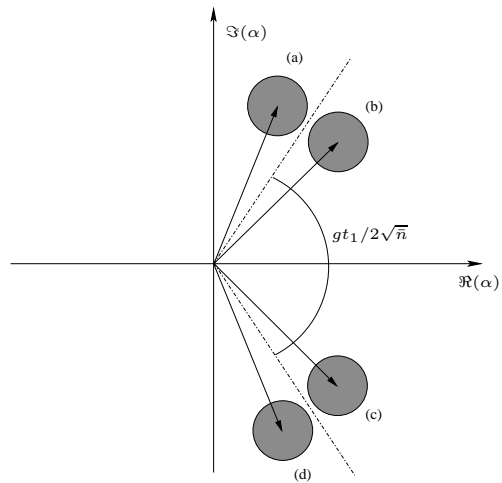}%
\end{picture}%
\setlength{\unitlength}{2072sp}%
\begingroup\makeatletter\ifx\SetFigFont\undefined%
\gdef\SetFigFont#1#2#3#4#5{%
  \reset@font\fontsize{#1}{#2pt}%
  \fontfamily{#3}\fontseries{#4}\fontshape{#5}%
  \selectfont}%
\fi\endgroup%
\begin{picture}(5667,5721)(2132,-6370)
\put(7501,-4018){\makebox(0,0)[lb]{\smash{{\SetFigFont{6}{7.2}{\familydefault}{\mddefault}{\updefault}{\color[rgb]{0,0,0}$\Re{(\alpha)}$}%
}}}}
\put(5986,-3436){\makebox(0,0)[lb]{\smash{{\SetFigFont{6}{7.2}{\familydefault}{\mddefault}{\updefault}{\color[rgb]{0,0,0}$gt_1/2\sqrt{\bar{n}}$}%
}}}}
\put(4715,-1018){\makebox(0,0)[lb]{\smash{{\SetFigFont{6}{7.2}{\familydefault}{\mddefault}{\updefault}{\color[rgb]{0,0,0}$\Im{(\alpha)}$}%
}}}}
\end{picture}%
 \caption{\label{fig:relaxation:one-jump} Quasi-coherent components generated after 
 one relaxation jump pictured in the Fresnel plane. Field quasi-coherent states are represented as
 uncertainty disks at the tip of their classical amplitudes. Components components (b) and (d) move clockwise
 whereas components (a) and (c) move counterclockwise.}
\end{figure}

When the first relaxation quantum jump occurs at time $t_{1}$, the qubit gets projected 
on the $|-\rangle$ state leaving the
cavity in a superposition of quasi-coherent states $|\psi_{\pm}(t_{1})\rangle$. The 
subsequent Rabi oscillation
signal then shows rapid oscillations of frequency $g\sqrt{\bar{n}}$ 
which correspond to the
superposition of the Rabi oscillation signals associated with the 
evolution of $|-\rangle\otimes |\psi_{\pm}(t_{1})\rangle$.
Therefore, the first relaxation jump leaves us with four quasi-coherent components in the Fresnel
plane as shown in fig. \ref{fig:relaxation:one-jump}:
\begin{equation}
\label{eq:relaxation:quantum-jump}
\sigma^-\,|\Psi^X_{\pm}(\theta)\rangle = 
\frac{e^{i\phi_{\pm,\theta}}}{\sqrt{2}}\,
\left(|\Psi^X_{+}(\theta)\rangle -|\Psi^X_{-}(-\theta)\rangle
\right)
\end{equation}
The Rabi oscillations starting right after the first jump collapse in time $g^{-1}$ due to the 
splitting of the pairs of counter-rotating quasi components in the Fresnel plane. Revivals
then occur when two of them recombine.

Assuming that $gt_{1}/2\sqrt{\bar{n}}<\pi$, the first recombination occurs
at time $2t_{1}$ and involves components (b) and (c) on fig. \ref{fig:relaxation:one-jump}).
Components (a) and (b) will recombine at time $t_{1}+4\pi\sqrt{\bar{n}}/g$ as
well as components (c) and (d). But in these three cases,
the rapid oscillations are simply washed out by the averaging over $t_{1}$ as in the pure
dephasing case. Only the recombination of components (a) and (d) will not be averaged to zero 
since it takes place at time $4\pi\sqrt{\bar{n}}/g$, independent of $t_{1}$.

A careful discussion of of all contributions is presented in appendix \ref{appendix:dynamics}. It
shows that keeping only the first term in the r.h.s. of
\eqref{eq:relaxation:quantum-jump} corresponds to
retaining only trajectories for rapid oscillations survive the averaging  over dates of relaxation jumps. 
Therefore, we are brought back to the
case of a single pair of trajectories within the Fresnel plane. But with each relaxation jump comes
not only a pure phase as in eq. (38) of \cite{Degio22} but a coefficient $e^{-i\phi_{m,\theta}}/\sqrt{2}$.
Following \cite{Degio22}, these coefficients can be resummed thus leading to the 
decoherence coefficient associated with relaxation in front of the coherence 
$|\Psi_{+}^X(t)\rangle\langle \Psi_{-}^X(t)|$:
\begin{equation}
\label{eq:relaxation:decoherence-coefficient}
\mathcal{F}^{(R)}_{+-}(t)=\exp{\left(\frac{\gamma_{1}}{2}\int (\frac{1}{2}\,e^{i\Delta\phi(\tau)}-1)\,d\tau\right)}
\end{equation}
where $(\Delta\phi)(\tau)=g\tau/2\sqrt{\bar{n}}$ denotes the angular separation in the
Fresnel plane of the quasi-component components of $|\Psi_{\pm}^X(\tau)\rangle$. 

As a final comment of this discussion, 
let us point out that this decoherence coefficient alone is not sufficient to keep track of the
complete state of the atoms + cavity system. More precise results could be obtained by perfoming 
a pertubative expansion in $\gamma_{1}/g$ keeping track of all contributions arising from
\eqref{eq:relaxation:quantum-jump}. Cavity losses and dephasing jumps can then be taken into account.
The resulting formula are appropriate in the weak relaxation regime but rapidely 
require numerical evaluation. 

\subsubsection{Dispersive regime}
 \label{sec:dissipation:relaxation:dispersive}

Being off resonance, the cavity cannot send back the qubit to the $|+\rangle$ state after a relaxation
quantum jump. As stressed before, since the states $|\pm\rangle\otimes |\alpha\rangle$ are pointer 
states with respect to cavity 
relaxation, the relaxation dynamics is most conveniently studied in term of these states. 

\medskip

The state $|-\rangle \otimes |\alpha\rangle$ are obviously left invariant by the relaxation
dynamics since $\sigma^-\ldotp\,|-\rangle \otimes |\alpha\rangle=0$. The state 
$|+\rangle\otimes|\alpha\rangle$ is sent on $|-\rangle\otimes|\alpha\rangle$ by the relaxation
jump. The probability for such a jump to occur between $t$ and $t+dt$ is $\gamma_{1}\,dt$, independent
of $\alpha$. With this in mind,  an exact solution for the dynamics in the dispersive regime,
taking into account all dissipative processes (relaxation of the cavity, qubit relaxation and dephasing)
can now be given for an initial state of the form $(A_{+}|+\rangle+A_{-}|-\rangle)\otimes |\alpha\rangle$.

\medskip

Following Ref. \cite{Gambetta:2006-1}, the qubit + cavity reduced density operator can be decomposed with
respect to the qubit state
$\rho(t)=
\sum_{\epsilon,\epsilon'}\rho_{\epsilon,\epsilon'}(t)\otimes 
|\epsilon\rangle\langle\epsilon'|$.
Note that $\rho_{-+}=\rho_{+-}^\dagger$ and $\mathrm{Tr}((\rho_{++}+\rho_{--})(t))=1$.
First of all, diagonal terms $\rho_{\epsilon,\epsilon}(t)$ are not affected by pure dephasing. 
Next, the $\rho_{+,+}(t)$ operator
only contains contributions coming from trajectories without any relaxation quantum jump whereas 
$\rho_{-,-}(t)$ is fed from trajectories that have had no quantum jump and trajectories having a single
jump at time $0\leq \tau\leq t$. This leads to:
\begin{eqnarray}
\label{eq:dispersive:solution:rho++}
\rho_{++}(t) & = & |A_{+}|^2e^{-\gamma_{1}t}\,|\alpha_{+}(t)\rangle\langle \alpha_{+}(t)|\\
\rho_{--}(t) & = & |A_{-}|^2\,|\alpha_{-}(t)\rangle\langle \alpha_{-}(t)|\nonumber \\
& + & \gamma_{1}e^{-\gamma_{1}t}\int_{0}^t
|\tilde{\alpha}_-(t,\tau)\rangle\langle\tilde{\alpha}_-(t,\tau)| \,d\tau
\label{eq:dispersive:solution:rho--}
\end{eqnarray}
where $\alpha_{\pm}(t)$ denotes the solution to \eqref{eq:pointer:cQED:dispersive:eqlambda} 
with corresponding sign and initial condition $\alpha_\pm(0)=\alpha$
and $\tilde{\alpha}_-(t,\tau)$ denotes 
the solution of \eqref{eq:pointer:cQED:dispersive:eqlambda} with $+$ sign from $0$ to $\tau$ and
then solution to \eqref{eq:pointer:cQED:dispersive:eqlambda} with $-$ sign.

Because coherence operators $\rho_{+-}(t)$ and $\rho_{-+}(t)$ involves
a $|-\rangle$ state, they only receive contributions from trajectories without any quantum
jump. Taking into account pure dephasing leads to:
\begin{eqnarray}
\rho_{+-}(t) & = & A_{+}A_{-}^*e^{-(\gamma_{\varphi}+\frac{\gamma_{1}}{2})t}
\mathcal{F}_{+-}(t)\nonumber \\
& \times & e^{i(\varphi_{+}-\varphi_{-})(t)}\,|\alpha_{+}(t)\rangle\langle\alpha_{-}(t)|\,.
\label{eq:dispersive:solution:rho+-}
\end{eqnarray}
These formula are nothing but generalizations of eqs. (5.14) to (5.18) of \cite{Gambetta:2006-1} that
take into account relaxation of the qubits and for a general driving of the cavity. 

Our derivation, not relying on the $P$-function
formalism, provides a simple view of the underlying quantum dynamics showing the importance
of pointer states with respect to to cavity losses and qubit pure dephasing. Following the
same line of reasoning, the reduced density operator for several qubits coupled to
a driven cavity all in the dispersive regime can be computed although it leads to more complicated
expressions.
 
 \subsection{Rabi oscillation signals}
 \label{sec:dynamics:Rabi-results}
 
 \subsubsection{General form of the signal}
 \label{sec:dynamics:Rabi-results:general}
 
Let us now use these results by computing 
the Rabi oscillation signal and its enveloppe assuming that the average photon number in the
 cavity remains equal to $\bar{n}$. The result for the Rabi oscillation signal $P(t)$ defined 
 as the probability of finding the qubit in the $|+\rangle$ state is given by:
\begin{equation}
\label{eq:Rabi:free:general}
P(t)=\frac{1}{2}\left(1+ \Re{\left( 
e^{-igt\sqrt{\bar{n}}/2+i\frac{gt}{2\sqrt{\bar{n}}}}
\mathcal{R}_{+-}(t)\,\mathcal{F}_{+-}(t)
\right)}\right)\,.
\end{equation}
where $\mathcal{R}_{+-}(t)=\langle \psi_{-}(t)|\psi_{+}(t)\rangle$ 
measures the overlap between the two Gea-Banacloche
quasi coherent component of the field.
Decoherence is contained in 
$\mathcal{F}_{+-}(t)$. Given these coefficients, 
the upper and lower enveloppes $P_{\pm}$ of the Rabi oscillation signals are given,
 in the case of a single qubit initially in the excited state, by \cite{Degio22}:
 \begin{equation}
\label{eq:enveloppes:N=1}
P_{\pm}(t) = \frac{1}{2}\left(1\pm
|\mathcal{R}_{+-}(t)\,\mathcal{F}_{+-}(t)|\right)\,.
\end{equation} 
Explicit expressions for the overlap and decoherence coefficients will now
be given for the case of free evolution of the cQED system (spontaneous revivals)
 and for the case of an echo experiment (induced revivals).
  
 \medskip
  
 \subsubsection{Spontaneous revivals}
 \label{sec:dynamics:Rabi-results:free-evol}
 
In the case of a 
free evolution (spontaneous revivals), the overlap factor $\mathcal{R}_{+-}(t)$ is given by:
\begin{equation}
\label{eq:free:overlap}
\mathcal{R}_{+-}(t)=e^{-\bar{n}}e^{igt\sqrt{\bar{n}}}
\sum_{k=0}^{+\infty} \frac{\bar{n}^k}{k!}\,e^{-igt\sqrt{k+1}}\,.
\end{equation}
Taking into account all decoherence sources, the decoherence factor 
$\mathcal{F}_{+-}(t)=e^{-d(t)+i\Theta(t)}$ is given by:
 \begin{eqnarray}
d(t) & = & \left(\kappa\bar{n} +\frac{\gamma_{1}+\gamma_{\varphi}}{2}\right)\,t\nonumber\\
& - & \frac{2\sqrt{\bar{n}}}{g}\left(\kappa\bar{n}+\frac{\gamma_{1}}{4}\right)\,
\sin{(\phi_{t})}
\label{eq:free:decay}
\end{eqnarray}
where $\phi_{t}=gt/2\sqrt{\bar{n}}$ and
\begin{equation}
\Theta(t) =  \frac{\gamma_{1}+4\kappa\bar{n}}{g}\,
\sin^2{\left(\frac{\phi_{t}}{2}\right)}\,.
\label{eq:free:phase}
\end{equation}
 Decoherence is mainly dominated by an exponential decay with rate 
 $\Gamma=\kappa\bar{n}+(\gamma_{\varphi}+\gamma_{1})/2$. The appearance of
 $\kappa\bar{n}$ reflects the photon emission rate enhancement by stimulated emission.
 The $1/2$ reduction of the relaxation rate comes from the averaging over classical Rabi
 oscillations of the qubit relaxation rate. To explain the $1/2$ reduction of the dephasing 
let us note that, for a standalone qubit, contributions from all 
 quantum history pile up to give 
 $\langle +|\rho(t)|+\rangle=\langle +|\rho(0)|+\rangle\, e^{-\gamma_{\varphi}t}$. 
 On the contrary, for a qubit coupled to a mesoscopic
 field in a cavity, only quantum histories without any quantum jump before the measurement
 time contribute to the Rabi oscillation signal. 
 
 \subsubsection{Induced revivals}
 \label{sec:dynamics:Rabi-results:echo}
 
 In the echo experiment with echo pulse performed at time $t_{\pi}$, the Rabi oscillation signal at time
 $t\leq t_{\pi}$ is determined by eqs. \eqref{eq:free:overlap},
 \eqref{eq:free:decay} and \eqref{eq:free:phase}. For $t\leq t\pi$, the overlap factor is given by
 $\mathcal{R}_{+-}(t_\pi,t)=\mathcal{R}_{+-}(2t_{\pi}-t)$ where the r.h.s. involves the free evolution
 overlap factor since the echo pulse reverses the dynamics of the qubit + cavity state \cite{Morigi:2002-1}.
 
 The decoherence factor
 $\mathcal{F}_{+-}(t_{\pi},t)=e^{-d(t_{\pi},t)+i\Theta(t_{\pi},t)}$ for $t\geq t_{\pi}$ is now given
 in terms of:
 \begin{eqnarray}
 d(t_{\pi},t) & = &
  \frac{2\sqrt{\bar{n}}}{g}\left(\kappa\bar{n}+\frac{\gamma_{1}}{4}
 \right)\,\lbrace\sin{\left(2\phi_{\pi}-\phi_{t}\right)}
 -2\sin{(\phi_{\pi})}\rbrace \nonumber\\
 & + &
   \left(\kappa\bar{n}+\frac{1}{2}(\gamma_{\varphi}+\gamma_{1})\right)\,t
 \end{eqnarray}
 where $\phi_{\pi}=gt_{\pi}/2\sqrt{\bar{n}}$. Its phase 
 is the sum of a contribution due to cavity relaxation (eq. (51) of \cite{Degio22}) and
 a contribution due to qubit relaxation:
 \begin{equation}
 \Theta_1(t_{\pi},t)=
 \frac{\gamma_{1}\sqrt{\bar{n}}}{g}\,
 \left(2\sin^2{\left(\frac{\phi_{\pi}}{2}\right)}
 -\sin^2{\left(\phi_{\pi}-\frac{\phi}{2}\right)}\right)
 \end{equation}
 Note that the dephasing exponential factor $e^{-\gamma_{\varphi}t/2}$ 
 is the same than in the free evolution since the echo pulse is not
 able to reverse the effect of a Markovian qubit dephasing. Besides the overlap coefficient 
 $\mathcal{R}_{+-}(t_{\pi},t)$, the echo pulse mainly affects the contribution arising from the accumulation
 of slow phases either associated to cavity relaxation or to qubit relaxation.

 \section{Discussion of the results}
 \label{sec:results}
 
 \subsection{Methods and parameters}
 \label{sec:results:method}
 
The Yale group reports of values of $g/2\pi$ between
$5.8$ and $100$~MHz \cite{Blais:2007-1}.
Low relaxation and dephasing rates have recently been
measured \cite{Wallraff:2005-1} at the magic point $n_{g}=1/2$: $\kappa/2\pi\simeq 0.6$~MHz,
$\gamma_{1}/2\pi\simeq 0.02$~MHz and
$\gamma_{\varphi}/2\pi\simeq 0.31$~Mhz but 
for a low value of the coupling\footnote{Our coupling energy $g$ corresponds to the vacuum Rabi
splitting and is denoted by $2g$ in the Yale group papers.}: 
$g/2\pi\simeq 11.6$~MHz (see row Circuit QED (1)). 
In a more recent experiment \cite{Schuster:2007-1}, 
a coupling of $g/2\pi\simeq 210$~MHz has been obtained with dissipation parameters
$\kappa/2\pi\simeq 0.25$~MHz, $\gamma_{1}/2\pi\simeq 1.8$~MHz and 
$\gamma_{\varphi}/2\pi\simeq 1$~Mhz (see row Circuit QED (2) in table \ref{table:1}).
The last row of table \ref{table:1} present choices of dissipation parameters compatible
with near future improvements of circuit QED devices. 
   
\begin{table}
\begin{ruledtabular}
\begin{tabular}{|c|c|c|c|} \hline
 Type of device & $g/\kappa$ &  $g/\gamma_{1}$ & $g/\gamma_{\varphi}$\\
\hline \hline 
Rydberg atoms (1) & $310$ & $10230$ & $\infty$ \\ 
Rydberg atoms (2) & $4300$ & $10230$ & $\infty$ \\ 
Circuit QED (1) & $19.4$  & $580$ & $40$\\ 
Circuit QED (2) & $840$ & $106$ & $215$ \\
Circuit QED (3) & $1400$ & $2000$ & $2000$ \\ \hline
\end{tabular}
\end{ruledtabular}
\caption{\label{table:1} 
Physical parameters for cQED experiments performed with Rydberg atoms (LKB group) or with superconducting
circuits (Yale group). Experimentally realized configurations appear on 
rows Rydberg atoms (1) \cite{Meunier:2005-1}, Circuit QED (2) \cite{Wallraff:2005-1}
and Circuit QED (2) \cite{Schuster:2007-1}.}
\end{table} 
 
 Analytical results presented in section \ref{sec:dynamics:Rabi-results} 
 have been compared to quantum Monte-Carlo
 simulations of the qubit + cavity evolution. For these simulations, the Adams-Blashford scheme of
 order four has been used to compute the evolution under the non-hermitian Hamiltonian between
 quantum jumps. All simulation results presented here have been obtained using an average over 10000
 trajectories. 
 
 Results for the free evolution of the qubit + cavity system will be first presented in section 
 \ref{sec:results:free} and
 for the echo experiment in section \ref{sec:results:echo}. 
 Consequences from these results are drawn in section \ref{sec:results:discussion}
 
 \subsection{Free evolution}
 \label{sec:results:free}
 
\begin{figure}
 \begin{center}
 \includegraphics[width=8.5cm]{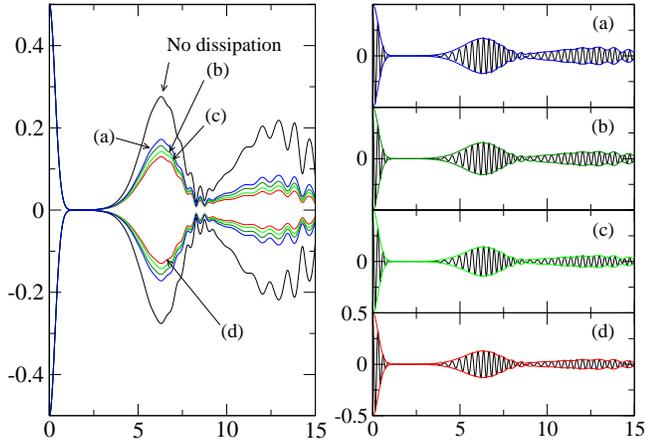}
 \end{center}
 \caption{\label{fig:free-evolution:1} (Color online)
 Influence of the various dissipation mechanisms on the spontaneous revivals for
 a single qubit in the presence of a $\bar{n}=10$ photon field. The graph on the left
 depicts the analytical enveloppes for the Rabi oscillation signal $S^z(t)=P_{m=1/2}(t)-1/2$ as a 
 function of $t/t_{R}$ for no dissipation (black lines), 
 (a) only cavity dissipation $g/\kappa=840$, (b) cavity and qubit relaxation $g/\gamma_{1}=106$,
 (c) cavity relaxation and qubit pure dephasing $g/\gamma_{\varphi}=210$ and 
 (d) all dissipation mechanisms present. The right part of the figure presents the associated Rabi
 oscillation signals obtained from quantum Monte-Carlo simulations (plain line) as well as the
 corresponding analytical enveloppes in cases (a) to (d).}
 \end{figure}

As a first step, it is interesting to see how our analytical model provides accurate predictions in
the presence of qubit relaxation and pure dephasing on top of cavity relaxation. Figure \ref{fig:free-evolution:1}
presents a comparison of the analytical enveloppes for the Rabi oscillation signal $S^z(t)=P_{m=1/2}(t)-1/2$
in the presence of (a) cavity relaxation alone, (b) cavity and qubit relaxation but no pure dephasing, (c) cavity
relaxation and pure dephasing and finally (d) including all dissipative mechanisms. The
dissipative parameters used for this comparison correspond to the Circuit QED (2) row of table
\ref{table:1}: $g/\kappa=840$, $g/\gamma_{1}=106$ and $g/\gamma_{\varphi}=210$.
With these parameters, the analytical model correctly describes the enveloppe of the Rabi oscillation signal.

\begin{figure}
 \begin{center}
\includegraphics[width=8.5cm]{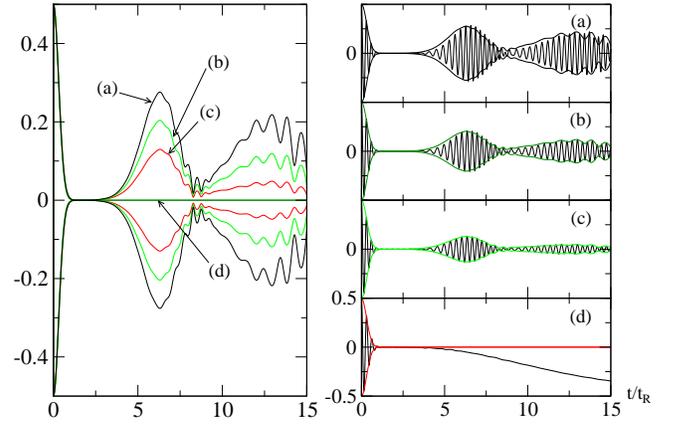}
 \end{center}
 \caption{\label{fig:free-evolution:2} (Color online)
 Spontaneous revivals for a single qubit in the presence of a $\bar{n}=10$ photon field for
 various dissipative parameters. The graph on the left
 depicts the analytical enveloppes for the Rabi oscillation signal $S^z(t)=P_{m=1/2}(t)-1/2$ as a 
 function of $t/t_{R}$ for (a) no dissipation, (b) Circuit QED (3) parameters,
 (c)  Circuit QED (2) parameters
 and (d) Circuit QED (1) parameters (see table \ref{table:1}). 
 The right part of the figure presents the associated Rabi
 oscillation signals obtained from quantum Monte-Carlo simulations (black lines) as well as the
 corresponding analytical enveloppes in cases (a) to (d).
 }
 \end{figure}

Figure \ref{fig:free-evolution:2} compares the Rabi oscillation signals for 
the experimental parameters corresponding to Circuit QED lines in table \ref{table:1}. 
In cases (a) to (c), our anaytical model predicts the upper and lower enveloppes for the Rabi oscillation
signal with good precision. But in case (d), only the initial collapse is correctly accounted for by the
analytical enveloppes. Our Monte-Carlo simulation shows a relaxation of $S^z(t)$ towards $-1/2$ at longer times
which is not described by our analytical model. This result is not surprising since for $g/\kappa=19.4$, at
$t=5t_R$, the average photon number in the cavity should have decayed from 10 to 2 because of
cavity relaxation alone. Therefore, at $t\sim 5t_R$, the qubit + cavity system is already out of the mesoscopic
regime and our model is not expected to be valid in this regime.

Finally, these numerical and analytical results show that
observing spontaneous Rabi oscillation revivals in circuit QED is not possible with
the Circuit QED (1) parameters of table \ref{table:1}.  
Only the recent improvements on the value of $g/\kappa$ (Circuit QED (2) parameters) open
the possibility for observing this phenomenon. 
 
\begin{figure}
\includegraphics[width=9cm]{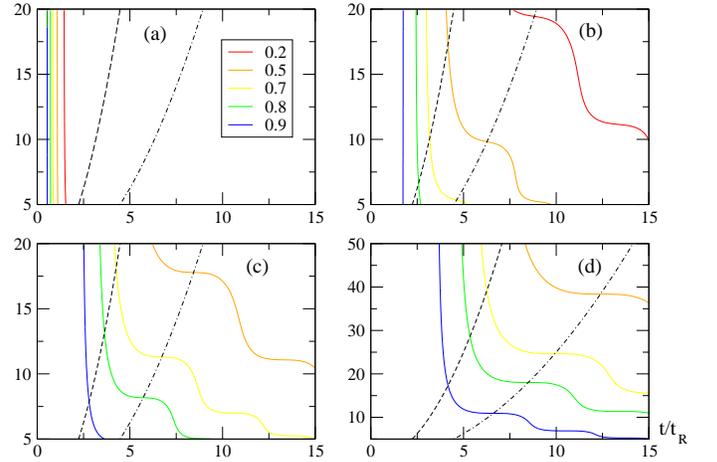}
\caption{\label{fig:free-evolution:3} (Color online)
Coutour plots of the effective decoherence coefficient $|\mathcal{F}_{+-}(t)|$ as a function of
$t/t_{R}$ and $\bar{n}$ ($\bar{n}\geq 5$ to ensure validity of the mesoscopic
approximation). The first three graphs represent the case of circuit QED devices: (a)
Circuit QED (1) parameters, Circuit QED (2) parameters and (c) Circuit QED (3) parameters. Graph
(d) depicts the corresponding information for Rydberg atom experiments with $g/\kappa=4310$
corresponding to a cavity with 14~ms damping time. In this case we extended the $\bar{n}$ range up to
50 photons. On each graph, the dashed line corresponds to the time of Schrödinger cat preparation
$t=t_R\sqrt{\bar{n}}$ and the dot-dashed line correspond to the maximum of the first spontaneous Rabi 
oscillation revival $t=2t_R\sqrt{\bar{n}}$.}
\end{figure}

The influence of photon number can be visualized through
the effective decoherence coefficient  $C(t,\bar{n})=|\mathcal{F}_{+-}(t)|$ 
as a function of $t/t_{R}=gt/2\pi$ and $\bar{n}$. It contains the effect
of the environment on the contrast of Rabi oscillation revivals. 
Contour plots $C(t,\bar{n})=0.9$, $0.8$, $0.7$, $0.5$ and $0.2$ are shown on
figure \ref{fig:free-evolution:3} for the Circuit QED parameter sets of table \ref{table:1}.
The fourth graph shows these coutours for the case of a Rydberg atom experiment
assuming a cavity damping time of 14~ms. 

At high photon number, the various coutour curves tend
to become vertical. This is easily understood by noticing that, at high photon number,
the cavity decoherence coefficient becomes independant of $\bar{n}$ and dominated the
decoherence process. In this regime, $C_\kappa(t)\simeq \exp{(-2\pi^3\,(g/\kappa)\,(t/t_R)^3)}$ and
the decoherence time scales typically as $\pi t_R(2g/\kappa)^{1/3}$.
Figure \ref{fig:free-evolution:3} shows that a coherent mesoscopic dynamics could
be observed in really good conditions in cases (c) and (d): $C(t,\bar{n})\gtrsim 0.8$
at  $t=t_R\sqrt{\bar{n}}$ corresponding to the generation of a mesoscopic field Schr\H{o}dinger
cat state involving up to 15 photons.

 \subsection{Echo experiment}
 \label{sec:results:echo}

\begin{figure}
\includegraphics[width=8cm]{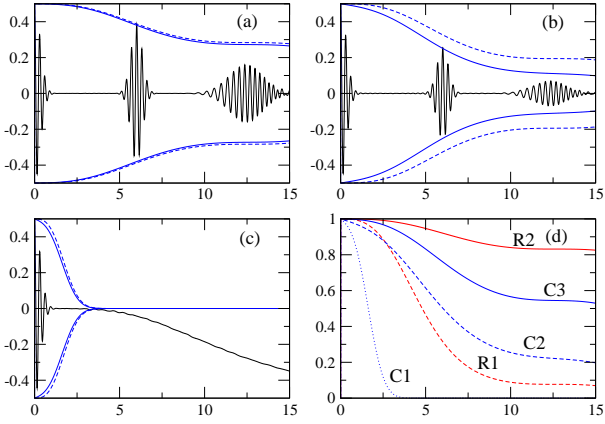}
\caption{\label{fig:echos:1} (Color online)
Induced revivals signal $S^z(t)$ for one qubit in the presence of a $\bar{n}=10$ 
photon field for various dissipative paramerters
as a function of $t/t_R$. Graphs (a) to (c) depict Rabi oscillation 
signals obtained from a quantum Monte-Carlo simulation
 for $t_\pi=3t_R$ (back curve). Analytical enveloppes showing the contrast at 
 the induced revival $|\mathcal{F}_{+-}(t/2,t)|$ taking into account all sources of 
 dissipation (full blue lines) or retaining only cavity relaxation (dashed blue lines) are also displayed.
Graph (d) depicts the contrast of the induced revival as a function of $t/t_R=2t_\pi/t_R$ 
using parameters of table \ref{table:1}.
The Circuit QED ($n$) parameter set corresponds to curve $Cn$ ($n=1$ to $3$) 
and the Rydberg atom ($m$) set parameter corresponds to curve $Rm$ ($m=1,2$).} 
\end{figure}

First of all, we have considered a circuit QED echo experiment
obtained with 10 photons initially and dissipative parameters
$g/\kappa=840$, $g/\gamma_{1}=106$ and $g/\gamma_{\varphi}=210$ (Circuit QED (2)
of table \ref{table:1}). Figure \ref{fig:echos:1} shows the echo signals for $t/t_{R}=3$ as well
as the analytical enveloppes obtained for $t=2t_{\pi}$ when the two quasi-coherent components
of the field recombine. 

\begin{figure}
\includegraphics[width=9cm]{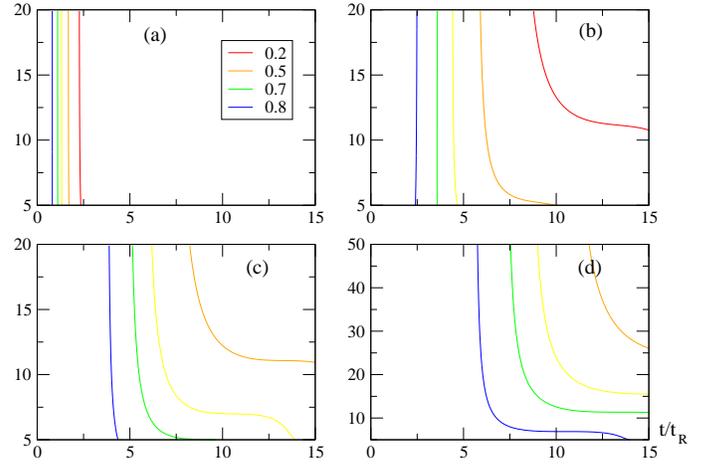}
\caption{\label{fig:echo:2} (Color online)
Coutour plots of the effective echo decoherence coefficient $|\mathcal{F}_{+-}(t/2,t)|$ as a function of
$t/t_{R}$ and $\bar{n}$ ($\bar{n}\geq 5$ to ensure validity of the mesoscopic
approximation). The first three graphs represent the case of circuit QED devices: (a)
Circuit QED (1) parameters, Circuit QED (2) parameters and (c) Circuit QED (3) parameters. Graph
(d) depicts the corresponding information for Rydberg atom experiments with $g/\kappa=4310$
corresponding to a cavity with 14~ms damping time. In this case we extended the $\bar{n}$ range up to
50 photons.}
\end{figure}

Because the echo experiment reverses exactly the dyamics of the qubit + cavity system, the 
reduction of the induced Rabi oscillation revivals contrast  is exactly given by the environmental
decoherence: $C_e(t,\bar{n})=|\mathcal{F}_{+-}(t/2,t)|$. Figure \ref{fig:echo:2} shows
coutour plots for $C_e(t,\bar{n})$ as a function ot $t/t_{R}$ and
$\bar{n}$ using the sets of parameters given in table \ref{table:1} for circuit QED devices.
As in the previous paragraph, the corresponding plot for Rydberg atom experiments with cavity damping
time of 14~ms are given.

 \subsection{Discussion of the results}
 \label{sec:results:discussion}
 
Within the context of circuit QED experiments, our results suggest that tests of quantum coherence
in the resonant mesoscopic regime could be performed using the latest generation of circuit QED
devices. Echo experiments could be performed with up to 10 photons and echo times up to
five vacuum Rabi periods. A more precise insight into the field dynamics could be obtained
by performing a field tomography giving access to either the $Q$ or the Wigner function of the field.
The $Q$ function probes the phase distribution of the field and its experimental determination would 
test the splitting of the field into two quasi-coherent components with opposite phases. 
Observing the induced revival then provides an experimental proof of the 
generation of a mesoscopic superposition superposition in the cavity \cite{Meunier:2005-1}. 

\medskip
 
On the other hand, Rydberg atom devices \cite{Raimond:2001-1} are characterized by very long 
atomic relaxation time (33~ms) and the absence of atomic dephasing. The main
limitation comes from cavity relaxation and time of flight. Up to 2005, the LKB team was able to
work with a $\kappa^{-1}\simeq 1$~ms 
cavity relaxation time corresponding to $g/\kappa\simeq 310$ 
\cite{Meunier:2005-1} (see row Rydberg atoms (1) in table \ref{table:1}). 
Cavity damping times up to 14~ms have been considered in our recent publication \cite{Degio22}
an are recalled in row Rydberg atoms (2) of table \ref{table:1}. Recently, the LKB
team reported on the realization of a ultra-high finesse cavity 
reaching $\kappa^{-1}\simeq 10$~Hz \cite{Kuhr:2006-1} corresponding to
$g/\kappa\simeq 31000$! With such performances, the only limitation is the atomic time of flight.
This is an
important limitation for the study of spontaneous revivals. In practice, only
partial spontaneous revivals arising from small atomic ensembles 
could be observed \cite{Degio22}. It is worth mentioning that circuit QED experiments do not
suffer from this limitation although their dissipative properties are still behind
the Rydberg atom ones. 

\section{Conclusion}
 
 Motivated by rapid experimental progresses of circuit QED devices, 
 we have studied the dissipative dynamics of one qubit resonantly coupled to a mesoscopic
 field in a cavity taking into account cavity relaxation as well as qubit relaxation and pure dephasing. 
 The resonant coupling between the qubit and the cavity produces an entangled state involving
 quasi-pointer states with respect to cavity relaxation. This very precise fact ultimately justifies in
 full generality the simple analytical model previously used 
 to describe the dynamics of Rydberg atom ensembles coupled to a high-Q cavity \cite{Degio22}. 
 Although quasi-pointer states naturally produced by the resonant interaction are not preserved by
 qubit relaxation and dephasing,
 we have obtained a simple description of the dynamics of the cavity + qubit system in the presence of 
 qubit dissipative processes. Simple 
 analytical expressions for the spontaneous and induced Rabi oscillation
 signals have been obtained and shown and to be 
 in good agreement with quantum Monte-Carlo simulations.
 
 \medskip
 
 Our study shows that these effects and the generation of mesoscopic Schrödinger cat states 
 could be within experimental reach in forthcoming circuit-QED experiments. However, our prediction
 may still be over optimistic since we did not
 attempt at modeling precisely the qubit readout.  
Thus, as done in Ref. \cite{Wallraff:2005-1} for the Ramsey fringes measurement,
the measurement beam can only be switched on after the qubit + cavity resonant interaction period, once
the qubit has been sent off resonance for dispersive readout. 
Obtaining more precise quantitative results suitable for describing
 an actual experiment will probably require a more 
 complete modeling of the experimental run, from preparation to measurement. The same remark is also
 relevant for obtaining precise predictions for the result of a cavity state tomography following for
 example a proposed protocol recently proposed in Ref. \cite{Melo:2006-1}. Ultimately, we may probably 
 have to rely on quantum Monte-Carlo simulations.
  
 \medskip 
 
 Other superconducting circuit for on-chip cavity QED have been proposed. In these
 proposals, the microstrip resonator is replaced by another oscillator such as an LC circuit or a biased 
 large Josepshon junction in the appropriate parameter regime. 
 Buisson and Hekking \cite{Buisson:2001-1} have proposed a circuit that couples
 a current biased dc-SQUID to a Cooper pair box. In principle, values of $g/2\pi$ of the 
 order of 100 to 200~MHz are within reach with such a device. However, modeling its dissipative
 dynamics requires taking into account the effect of low frequency noises which act on the qubit through low
 frequency dephasing and on 
 the resonator through slow fluctuations of its resonant frequency. These effects are easy to account for using our
 formalism and a quasi-static approximation. But the resonator is also coupled to high frequency current 
 noise through the SQUID \cite{Claudon:2006-1}. Work is in progress to describe the effect of this coupling
 on the Buisson-Hekking cavity-QED circuit.

\begin{acknowledgements}
This work was supported in part by the Quantum Condensed Matter Visitors Program of Boston University.
M. Clusel and J.-M. Raimond are thanked for their careful reading of the manuscript.
\end{acknowledgements}  

\appendix
\section{Non unitary evolution between quantum jumps}
\label{appendix:non-hermitian}

The non hermitian Hamiltonian describing the evolution between two quantum jumps
is:
\begin{equation}
\label{eq:non-hermitian:effective-hamiltonian}
\hbar^{-1}H=\hbar^{-1}H_{JC}-\frac{i\gamma_{1}}{2}\,\frac{1+\sigma^z}{2}-\frac{\kappa}{2}\,N 
-i\frac{\gamma_{\varphi}}{4}
\end{equation}
Obviously, the pure dephasing part leads to an trivial overall $e^{-\gamma_{\varphi}t/2}$ factor.
In the mesoscopic regime,
the resulting Hamiltonian can then be rewritten in terms of the $su(2)$ generators $\mathcal{J}^a$ suitable
for describing the dynamics of generalized Gea-Banacloche 
states \eqref{eq:generalized-GBstates} \cite{Degio22}:
\begin{eqnarray}
\label{eq:non-hermitian:effective-hamiltonian:final}
\hbar^{-1}H_{\mathrm{eff}} & = &  g\sqrt{p+1}\,\mathcal{J}^x
-\frac{i\,\kappa}{2}\,p\nonumber \\
 & - & \frac{i\,(\gamma_{1}-\kappa)}{2}\,\mathcal{J}^z
 -\frac{i}{4}(\gamma_{1}+\kappa)\,.
\end{eqnarray}
This Hamiltonian is of the same form as the one discussed in appendix~D 
of \cite{Degio22}. The resulting
dynamics can then be inferred straightforwardly. 
In \cite{Degio22}, since $g\gg \kappa$, 
the backaction of cavity relaxation on the qubit has been neglected and since $g\ll \gamma_{1}$,
the term proportional to $\gamma_{1}\mathcal{J}^z$ shall also be dropped out.
Finally, starting from a state of the form
$$|\Psi(0)\rangle=e^{-\bar{n}/2}\sum_{p=0}^\infty
\frac{\bar{n}^{p/2}}{\sqrt{p!}}\,|\Psi_{p}(0)\rangle
$$
and assuming that $\gamma t\ll 1$, so that the average photon number in the cavity
remains constant between $0$ and $t$, we get:
\begin{equation}
\label{eq:non-hermitian:GB-state-evolution}
|\Psi_{{\mathrm{n.j.}}}(t)\rangle = e^{-\frac{(\gamma_{\varphi}+\gamma_{1})t}{4}}
e^{-\frac{\kappa t}{2}}
e^{-\bar{n}}\sum_{p=0}^\infty \frac{\bar{n}^{p/2}}{\sqrt{p!}}
e^{-igt\sqrt{p+1}\mathcal{J}^x}\,|\Psi_{p}(0)\rangle\,.
\end{equation}
The exponentially decaying 
prefactors are due to the non-hermiticity of the Hamiltonian. They reflect the decaying weight in time
of quantum trajectories without any quantum jump between $0$ and $t$. The apparently surprising
appearance of $\gamma_{1}/4$ instead of $\gamma_{1}/2$ for relaxation quantum jumps corresponds to 
time averaging to $1/2$ of the qubit populations induced by the strong coupling with the cavity. 
Physically, we assume that the number of photons is so high and relaxation so weak that the cavity
mode keeps driving the qubit with the same intensity.
Note that the resulting effective dynamics correspond to an effective Poissonian statistics of relaxation quantum
jumps. But when the qubit state is close to $|-\rangle$, the probability rate for such
a jump is much smaller than when it is close to $|+\rangle$. Indeed
deviations from the effective Poissonian statistics average out since 
dissipative processes take place over a much longer time
scale than rapid oscillations induced by the resonant qubit/cavity coupling. 

 \section{Effects of dissipation on the resonant cQED system}
 \label{appendix:dynamics}
 
 In this section, the evolution of generalized Gea-Banacloche states \eqref{eq:generalized-GBstates}
under the stochastic dynamics involving qubit relaxation and dephasing is discussed.
We first derive the effective 
dynamics resulting from relaxation and pure dephasing quantum jumps.
The case of pure dephasing 
can then be treated exactly using renewal theory. A formal solution including relaxation
in the low dissipation, large photon number limit is provided and used to extract the dominant
contributions to the Rabi oscillation signal. This appendix provides a firm basis for the
disscusions of sections \ref{sec:dissipation:dephasing} and 
\ref{sec:dissipation:relaxation}.
 
 \subsection{Stochastic dynamics}
 \label{appendix:dynamics:resonant} 

 \subsubsection{Evolution of the coherences between generalized Gea-Banacloche states}
 \label{appendix:dynamics:resonant:general}

In order to compute the evolution of the qubit + cavity state, it is convenient to look 
at the evolution of a coherence between a given pair of generalized Gea-Banacloche states
with respective indices $(m_{+},m_{-})$.  Such a coherence is initially of the
form 
\begin{equation}
\label{eq:dynamics:coherence:definition}
\Pi_{m_{+},m_{-}}(0)=e^{-\bar{n}}\sum_{p_{+},p_{-}}
\frac{\bar{n}^{(p_{+}+p_{-})/2}}{\sqrt{p_{+}!\,p_{-}!}}\,
 |X_{m_{+}}^{(p_{+})}\rangle\langle X_{m_{-}}^{(p_{-})}|\,.
\end{equation}
As shown in appendix \ref{appendix:non-hermitian}, in the weak dissipation
limit and in the mesoscopic regime, 
the evolution between quantum jumps is exactly the same as in the dissipationless case apart from the normalization
prefactor that reflects the weight of quantum trajectories not presenting any quantum jump in the considered
time interval. Moreoever, in the following, we shall restrict ourselved to
short times ($\gamma t\ll 1$) so that
the average photon number in the cavity remains constant. 
The effect of various quantum jumps is described by:
\begin{eqnarray}
a\ldotp\,|X_{m}^{(p)}\rangle & = & \sqrt{p+1/2-m}\,|X_{m}^{(p)}\rangle 
\label{eq:appendix:jump:raw:photons}\\
\sigma^-\ldotp\,|X_{\pm }^{(p)}\rangle & = & 
(|X_{+}^{(p-1)}\rangle - |X^{(p-1)}_{-}\rangle)/2
  \label{eq:appendix:jump:raw:relaxation}\\
\sigma^z\ldotp \,|X_{m}^{(p)}\rangle & = & |X_{-m}^{(p)}\rangle\,.
 \label{eq:appendix:jump:raw:dephasing}
\end{eqnarray}
In principle, we should follow keep track of all
phases in each subspace generated by $|X_{\pm}^{(p)}\rangle$. It was proved in Ref. \cite{Degio22} that
for cavity relaxation,
this exact dynamics could be further simplified so that the resulting stochastic dynamics be expressed
only in terms of the generalized Gea-Banacloche 
states \eqref{eq:generalized-GBstates}. A similar route will be followed here and
recasting eqs. \eqref{eq:appendix:jump:raw:photons} to 
\eqref{eq:appendix:jump:raw:dephasing} in terms of generalized
Gea-Banacloche states using the mesoscopic approximation leads to:
 \begin{eqnarray}
 \label{eq:appendix:jump:photons}
 a\ldotp |\Psi_{m}^X(\theta)\rangle & = & \sqrt{\bar{n}}\,e^{i\phi_{m,\theta}}\,
 |\Psi^X_{m}(\theta)\rangle\\
  \label{eq:appendix:jump:relaxation}
 \sigma^-\ldotp |\Psi_{m}^X(\theta)\rangle & = & 
\frac{e^{i\phi_{m,\theta}}}{\sqrt{2}}\,
\left(|\Psi^X_{+}(\theta)\rangle -|\Psi^X_{-}(-\theta)\rangle
\right)\\
 \label{eq:appendix:jump:dephasing}
 (i\sigma^z)\ldotp |\Psi_{m}^X(\theta)\rangle & = & |\Psi^X_{-m}(-\theta)\rangle\,.
 \end{eqnarray}
 
 \subsubsection{Trajectories of generalized Gea-Banacloche states} 
  \label{appendix:dynamics:resonant:trajectories}
 
\begin{figure}
 \begin{center}
\begin{picture}(0,0)%
\epsfig{file=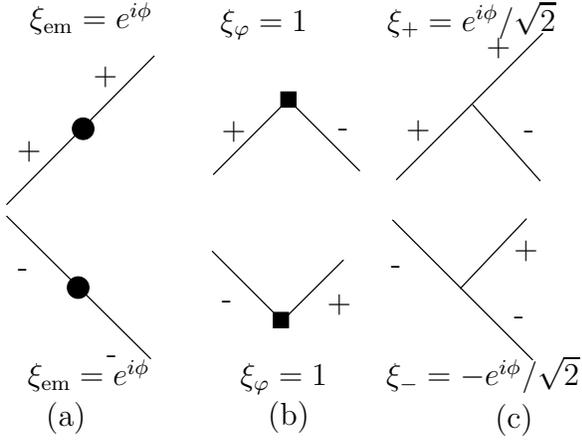}%
\end{picture}%
\setlength{\unitlength}{4144sp}%
\begingroup\makeatletter\ifx\SetFigFont\undefined%
\gdef\SetFigFont#1#2#3#4#5{%
  \reset@font\fontsize{#1}{#2pt}%
  \fontfamily{#3}\fontseries{#4}\fontshape{#5}%
  \selectfont}%
\fi\endgroup%
\begin{picture}(3233,2603)(2397,-4735)
\put(4689,-2303){\makebox(0,0)[lb]{\smash{{\SetFigFont{12}{14.4}{\familydefault}{\mddefault}{\updefault}{\color[rgb]{0,0,0}$\xi_{+}=e^{i\phi}/\sqrt{2}$}%
}}}}
\put(2478,-3774){\makebox(0,0)[lb]{\smash{{\SetFigFont{12}{14.4}{\familydefault}{\mddefault}{\updefault}{\color[rgb]{0,0,0}-}%
}}}}
\put(3010,-4292){\makebox(0,0)[lb]{\smash{{\SetFigFont{12}{14.4}{\familydefault}{\mddefault}{\updefault}{\color[rgb]{0,0,0}-}%
}}}}
\put(2474,-3073){\makebox(0,0)[lb]{\smash{{\SetFigFont{12}{14.4}{\familydefault}{\mddefault}{\updefault}{\color[rgb]{0,0,0}+}%
}}}}
\put(2931,-2630){\makebox(0,0)[lb]{\smash{{\SetFigFont{12}{14.4}{\familydefault}{\mddefault}{\updefault}{\color[rgb]{0,0,0}+}%
}}}}
\put(3698,-3968){\makebox(0,0)[lb]{\smash{{\SetFigFont{12}{14.4}{\familydefault}{\mddefault}{\updefault}{\color[rgb]{0,0,0}-}%
}}}}
\put(4329,-3991){\makebox(0,0)[lb]{\smash{{\SetFigFont{12}{14.4}{\familydefault}{\mddefault}{\updefault}{\color[rgb]{0,0,0}+}%
}}}}
\put(3698,-2956){\makebox(0,0)[lb]{\smash{{\SetFigFont{12}{14.4}{\familydefault}{\mddefault}{\updefault}{\color[rgb]{0,0,0}+}%
}}}}
\put(4395,-2934){\makebox(0,0)[lb]{\smash{{\SetFigFont{12}{14.4}{\familydefault}{\mddefault}{\updefault}{\color[rgb]{0,0,0}-}%
}}}}
\put(4801,-2949){\makebox(0,0)[lb]{\smash{{\SetFigFont{12}{14.4}{\familydefault}{\mddefault}{\updefault}{\color[rgb]{0,0,0}+}%
}}}}
\put(5506,-2949){\makebox(0,0)[lb]{\smash{{\SetFigFont{12}{14.4}{\familydefault}{\mddefault}{\updefault}{\color[rgb]{0,0,0}-}%
}}}}
\put(5281,-2454){\makebox(0,0)[lb]{\smash{{\SetFigFont{12}{14.4}{\familydefault}{\mddefault}{\updefault}{\color[rgb]{0,0,0}+}%
}}}}
\put(5446,-3668){\makebox(0,0)[lb]{\smash{{\SetFigFont{12}{14.4}{\familydefault}{\mddefault}{\updefault}{\color[rgb]{0,0,0}+}%
}}}}
\put(5446,-4059){\makebox(0,0)[lb]{\smash{{\SetFigFont{12}{14.4}{\familydefault}{\mddefault}{\updefault}{\color[rgb]{0,0,0}-}%
}}}}
\put(4711,-3759){\makebox(0,0)[lb]{\smash{{\SetFigFont{12}{14.4}{\familydefault}{\mddefault}{\updefault}{\color[rgb]{0,0,0}-}%
}}}}
\put(2536,-4426){\makebox(0,0)[lb]{\smash{{\SetFigFont{12}{14.4}{\familydefault}{\mddefault}{\updefault}{\color[rgb]{0,0,0}$\xi_{\mathrm{em}}=e^{i\phi}$}%
}}}}
\put(3810,-4426){\makebox(0,0)[lb]{\smash{{\SetFigFont{12}{14.4}{\familydefault}{\mddefault}{\updefault}{\color[rgb]{0,0,0}$\xi_\varphi=1$}%
}}}}
\put(5334,-4681){\makebox(0,0)[lb]{\smash{{\SetFigFont{12}{14.4}{\familydefault}{\mddefault}{\updefault}{\color[rgb]{0,0,0}(c)}%
}}}}
\put(3976,-4666){\makebox(0,0)[lb]{\smash{{\SetFigFont{12}{14.4}{\familydefault}{\mddefault}{\updefault}{\color[rgb]{0,0,0}(b)}%
}}}}
\put(2648,-4667){\makebox(0,0)[lb]{\smash{{\SetFigFont{12}{14.4}{\familydefault}{\mddefault}{\updefault}{\color[rgb]{0,0,0}(a)}%
}}}}
\put(3691,-2304){\makebox(0,0)[lb]{\smash{{\SetFigFont{12}{14.4}{\familydefault}{\mddefault}{\updefault}{\color[rgb]{0,0,0}$\xi_\varphi=1$}%
}}}}
\put(2551,-2288){\makebox(0,0)[lb]{\smash{{\SetFigFont{12}{14.4}{\familydefault}{\mddefault}{\updefault}{\color[rgb]{0,0,0}$\xi_{\mathrm{em}}=e^{i\phi}$}%
}}}}
\put(4682,-4426){\makebox(0,0)[lb]{\smash{{\SetFigFont{12}{14.4}{\familydefault}{\mddefault}{\updefault}{\color[rgb]{0,0,0}$\xi_{-}=-e^{i\phi}/\sqrt{2}$}%
}}}}
\end{picture}%
 \end{center}
 \caption{\label{fig:dynamics-diagrams} Tree structure associated with a stochastic trajectory with two
 photon emission, two pure dephasing quantum jumps and two relaxation jumps: (a) Vertices associated with the loss of a photon from the cavity
 (b) Vertices associated with a $\pi$ rotation of the qubit along the $z$§axis (pure dephasing) (c) Vertices associated with 
 relaxation of the qubit in its ground state (relaxation). 
} 
 \end{figure}

\begin{figure}
 \begin{center}
 \includegraphics{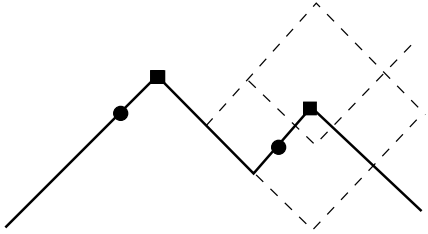}
 \end{center}
 \caption{\label{fig:typical-trajectory} The diagrammatic representation of a typical quantum stochastic 
 trajectory $\frak{T}$ showing two photon emissions from the cavity, two relaxations and two pure dephasing quantum 
 jumps. Time would correspond to an horizontal axis whereas the vertical axis is
 the angle of the quasi generalized Gea-Banacloche state in the Fresnel plane. 
 A single path $\mathcal{T}$ along $\frak{T}$ is singularized (bold line).}
 \end{figure}

 A sequence of quantum jumps of the above types will generate a tree structure whose
 vertices are associated with the branching expressed in eq. \eqref{eq:appendix:jump:relaxation}. Vertices
 associated with photons are located on straight edges of the tree 
 and vertices associated with pure dephasing jumps
 are associated with direction reversals (see fig. \ref{fig:dynamics-diagrams}). 
 The state obtained from a given stochastic trajectory $\frak{T}$ is a superposition of $2^{R_{\frak{T}}}$
 terms corresponding to all the paths along this tree structure where $R_{\frak{T}}$ denotes the number
 of relaxation events along $\frak{T}$ (see
  fig. \ref{fig:typical-trajectory} for an example where $\frak{T}$ has two relaxation jumps, two
 dephasing jumps and two photon jumps). 
 Note that each quantum jump $\alpha$ introduces a slow phase factor $\xi_{\alpha}$
 depending on its type and its occurence time (see fig. \ref{fig:dynamics-diagrams}). 
 Finally, the state $ |\Psi_{\frak{T}}(t)\rangle$ associated
 with the quantum trajectory $\frak{T}$ has the form:
 \begin{equation}
 |\Psi_{\frak{T}}(t)\rangle =
 \sum_{\mathcal{T}\in \mathrm{Path}(\frak{T})} 
 \left(\prod_{0\leq t_{\alpha}\leq t}\xi_{\alpha}\right)\,
 |\Psi_{m'_{t}}(\theta_{\mathcal{T}}(t))\rangle \,
 \end{equation}
where the sum in the r.h.s is performed over all paths $\mathcal{T}$ going
 downward the tree structure generated by quantum jumps along $\frak{T}$ (black 
 line in fig. \ref{fig:typical-trajectory}).
 In this expression, $m'_{t}=(-1)^{J[\mathcal{T}]}m$ depends on the number $J[\mathcal{T}]$ of
 direction changes along the path $\mathcal{T}$. The angle
  $\theta_{\mathcal{T}}(t)$ is related to the
 Fresnel angle $\phi_{\mathcal{T}}(t)$ by $\phi_{\mathcal{T}}(t)=
 -m'_t\theta_{\mathcal{T}}(t)/2\sqrt{\bar{n}}$. 
 
\subsubsection{Summing over stochastic trajectories} 
 \label{appendix:dynamics:resonant:resumming}

With this simple description of the stochastic dynamics, the evolution of the coherence
\eqref{eq:dynamics:coherence:definition} is of the form:
\begin{equation}
\label{eq:dynamics:coherence:evolution}
\Pi_{m_{+},m_{-}}(t) =\sum_{m'_{+},m'_{-}}\Pi^{(m'_{+},m'_{-})}_{m_{+},m_{-}}(t)
\end{equation}
where $\Pi^{(m'_{+},m'_{-})}_{m_{+},m_{-}}(t)$ only involves coherence between the 
$|X_{m'_{+}}^{(p)}\rangle$ and $|X_{m'_{-}}^{(p)}\rangle$ states with $m'_{\epsilon}=
\pm m_{\epsilon}$. Explicitely
  \begin{eqnarray}
 \Pi^{(m'_{+},m'_{-})}_{m_{+},m_{-}}(t) & = & e^{-\bar{n}}
  \sum_{p_{+},p_{-}}
  \frac{\bar{n}^{(p_{+}+p_{-})/2}}{\sqrt{p_{+}!\,p_{-}!}}\nonumber \\
  & \times & 
  \mathcal{D}^{(p_{+},p_{-})}_{m'_{+},m'_{-}}(t)
  |X_{m'_{+}}^{(p_{+})}\rangle\langle X_{m'_{-}}^{(p_{-})}|
  \end{eqnarray}
  where the coefficient 
   $\mathcal{D}^{(p_{+},p_{-})}_{m'_{+},m'_{-}}(t)$
(which will be denoted by $\mathcal{D}_{p_{+},p_{-}}(t)$ to shorten notation) is given by a path
  integral over all pairs of paths $\mathcal{T}_{\pm}$ downwards all possible pairs of
  quantum trajectories $\frak{T}_{\pm}$ of the appropriate phase:
  \begin{eqnarray}
  \label{eq:decoherence:double-path-integral}
  \mathcal{D}_{p_{+},p_{-}}(t) & = & \int 
  \mathcal{D}[\mathcal{T_{+}},\mathcal{T_{-}}]\,
  S_{t}[\mathcal{T}_{+}]\times S_{t}[\mathcal{T}_{-}]^*\nonumber \\
  & \times &
e^{2i\bar{n}(\sqrt{p_{+}+1}\phi_{\mathcal{T}_+}(t)
-\sqrt{p_{-}+1}\phi_{\mathcal{T}_-}(t))}\,.
  \end{eqnarray}
  Here $S_{t}[\mathcal{T}_{\pm}]$ are the products taken along $\frak{T}_{\pm}$
  of slow phases appearing in
  eqs. \eqref{eq:appendix:jump:photons} and \eqref{eq:appendix:jump:relaxation}:
  \begin{equation}
  \label{eq:decoherence:slow-phases}
 S_{t}[\mathcal{T}_{\pm}]=\prod_{0\leq t_{\alpha_{\pm}}\leq t}\xi_{\alpha_{\pm}}\,.
 \end{equation}

 \subsubsection{Atomic observables}
 \label{appendix:dynamics:atomic-observables}
 
Let us assume that the atoms + cavity reduced density operator at time $t$ is of the form given by the r.h.s of
\eqref{eq:dynamics:coherence:evolution}. The Rabi oscillation signal is completely contained in
the qubit polarization:
\begin{equation}
\label{eq:qubit-observables:population}
\langle \sigma^z(t)\rangle =e^{-\bar{n}}\sum_{p}\frac{\bar{n}^p}{p!}\,\frac{1}{2}\left(
\mathcal{D}_{+-}^{(p,p)}(t)+\mathcal{D}_{-+}^{(p,p)}(t)\right)
\end{equation}
and clearly involves the evolution of coherences between $|\Psi^X_{+}\rangle$
and $|\Psi^X_{-}\rangle$. Note however that it does not probe the coherence between subspaces associated
with different values of $p$. The qubit coherence is given by:
\begin{eqnarray}
\langle \sigma^+(t)\rangle  & = & e^{-\bar{n}}\sum_{p}\frac{\bar{n}^p}{2\,p!}\,\frac{1}{2}\left(
\mathcal{D}_{++}^{(p,p+1)}(t)-\mathcal{D}_{--}^{(p,p+1)}(t)\right. \nonumber \\
& + & \left. 
\mathcal{D}_{+-}^{(p,p+1)}(t)-\mathcal{D}_{-+}^{(p,p+1)}(t)\right)
\label{eq:qubit-observables:coherences}
\end{eqnarray}
and is sensitive to the coherence between adjacent multiplets.

\subsection{Decoherence coefficients}
 \label{appendix:dynamics:details}

The main problem is now to compute all decoherence coefficients 
 $\mathcal{D}_{p_{+},p_{-}}(t)$ (which also depends on $m_{\pm}$ and $m'_{\pm}$). 
 As we shall see, this cannot be done exactly 
 but a careful analysis of these sums in the mesoscopic regime will show us which trajectories
 contribute to the Rabi oscillation signal.
 
  \medskip
 
Generically, the decoherence coefficient $\mathcal{D}_{p_{+},p_{-}}(t)$ contains 
slow phases arising from the phase factors associated with photon emissions and qubit relaxations but also
a rapid random phase $e^{2i\bar{n} \lambda_{p_{+},p_{-}} (\phi_{\mathcal{T_{+}}}-
\phi_{\mathcal{T_{-}}})(t)}$
arising from difference of the eigenenergies of the states $|X^{(p_{\pm})}_{m_{\pm}}\rangle$. Note
that in the absence of qubit relaxation and dephasing \cite{Degio22}, this rapid phase is not random and
therefore, decoherence is completely caused by the slow phases associated with photon losses of the cavity.
Here, relaxation and dephasing quantum jumps of the qubit lead to 
the decay of $\mathcal{D}_{p_{+},p_{-}}(t)$. Indeed, in the mesoscopic
regime, averaging of this rapid phase dominates the decay of the decoherence coefficient 
$\mathcal{D}_{p_{+},p_{-}}(t)$. To confirm this, 
let us consider the coefficient $D_{p_{+},p_{-}}(t)$ defined by forgetting slow phases
in $\mathcal{D}_{p_{+},p_{-}}(t)$:
\begin{equation}
D_{p_{+},p_{-}}(t)=\int \mathcal{D}[\mathcal{T_{+}},\mathcal{T_{-}}]\,
e^{2i\bar{n}(\sqrt{p_{+}+1}\phi_{\mathcal{T}_+}(t)
-\sqrt{p_{-}+1}\phi_{\mathcal{T}_-}(t))}\,.
\end{equation} 
As we shall see now, this expression can be studied analytically by adapting renewal theory techniques already
used to compute decoherence coefficients \cite{Degio20}. 

\subsubsection{Waiting time probabilities}
 \label{appendix:dynamics:details:waiting-times}

The effective statistics of quantum jumps 
is Poissonian (see appendix \ref{appendix:non-hermitian} for relaxation) and statistics of jumps
of different types are statistically uncorrelated.  Equivalently, the probability
distribution for waiting times between two jumps of type $j$ is exponential: 
$\psi_{j}(\tau)=\gamma_{j}\,e^{-\gamma_{j}\tau}$ where $\gamma_{\mathrm{cav}}=\bar{n}\kappa$,
$\gamma_{\mathrm{deph}}=\gamma_{\varphi}/2$ and $\gamma_{\mathrm{relax}}=\gamma_{1}/2$. 
This exponential factor is taken into account through the norm of the states 
\eqref{eq:non-hermitian:GB-state-evolution}. 
 
\subsubsection{Pure dephasing}
 \label{appendix:dynamics:details:dephasing}

 In this case, there are no slow phases and no branching. No approximation has been made beyond the
 mesoscopic approximation on the dissipationless problem. Hence we expect the following results to
 be valid even in the regime of strong dephasing $g\lesssim \gamma_{\varphi}$.
 
 \medskip
 
 The decoherence coefficients
 $\mathcal{D}^{(p_{+},p_{-})}_{m'_{+},m'_{-}}(t)$
are given by averages over a telegraphic noise:
 \begin{equation}
 \label{eq:decoherence:telegraphic-both}
 \mathcal{D}^{\mathrm{(e/o)}}_{\lambda}(t)
 =\langle e^{i\lambda \int_{0}^tX(\tau)\,d\tau}
 \rangle_{\mathrm{e}/\mathrm{o}}\,.
 \end{equation}
 Here $\lambda=g(m_{+}\sqrt{p_{+}+1}-m_{-}\sqrt{p_{-}+1})$ and $X(\tau)=\pm 1$ is a telegraphic
 noise characterized by the probability distribution of waiting times between two switching 
 $\psi(\tau)=(\gamma_{\varphi}/2)e^{-\gamma_{\varphi}\tau/2}$ and the initial condition $X(0)=1$.
 The $\mathrm{e}$ (respectively $\mathrm{o}$) symbol specifies that the average is performed over noise
 histories with an even (respectively odd) number of switchings between $0$ and $t$ ($m'_{\pm}=m_{\pm}$
 for the $\mathrm{e}$ case and $m'_{\pm}=-m_{\pm}$ in the $\mathrm{o}$ case).
 
 Renewal theory provides an elegant analytical solution for 
 the decoherence coefficients \eqref{eq:decoherence:telegraphic-both}. 
Introducing $\Pi_{0}(t)=\int_{t}^\infty\psi(\tau)\,d\tau$ and proceeding along 
the lines of  \cite{Degio21},
the Laplace transforms of the $\mathcal{D}_{\lambda}^{\mathrm{(e/o)}}$ is obtained as:
 \begin{eqnarray}
 L[\mathcal{D}_{\lambda}^{\mathrm{(e)}}](s) & = & 
 \frac{L[\Pi_{0}](s+i\lambda)}{1-L[\psi](s-i\lambda)L[\psi](s+i\lambda)}\\
 L[\mathcal{D}_{\lambda}^{\mathrm{(o)}}](s) & = & 
 \frac{L[\Pi_{\lambda}](s)}{1-L[\psi](s-i\lambda)L[\psi](s+i\lambda)}\\
 L[\Pi_{\lambda}](s) & = & \frac{L[\psi](s-i\lambda)\,(1-L[\psi](s+i\lambda))}{s+i\lambda}
 \end{eqnarray}
 Performing the inverse Laplace transform gives the explicit time
 dependence. The "odd" coefficient is thus given by:
 \begin{equation}
 \label{eq:decoherence:telegraphic-even:exact}
 \mathcal{D}^{\mathrm{(o)}}(t)  = \frac{\gamma_{\varphi}}{2(s_{+}-s_{-})}\,\left(e^{s_{+}t}-e^{s_{-}t}\right)
 \end{equation}
 where
 \begin{equation}
 s_{\pm}=-\frac{\gamma_{\varphi}}{2}\pm i\sqrt{\lambda^2-\gamma_{\varphi}^2/4}\,.
 \end{equation}
 The "even" coefficient has a somehow more involved expression:
 \begin{eqnarray}
 \mathcal{D}^{\mathrm{(e)}}_{\lambda}(t) & = & \frac{1}{2}\left(
 1-\frac{i\lambda}{s_{+}-s_{-}}\right)\,e^{s_{+}t}\nonumber \\
 & + & \frac{1}{2}\left(
 1+\frac{i\lambda}{s_{+}-s_{-}}\right)\,e^{s_{+}t}\,.
 \label{eq:decoherence:telegraphic-odd:exact}
 \end{eqnarray}
 Depending on the dimensionless coupling $\tilde{g}=2\lambda/\gamma_{\varphi}$, two distinct regimes
 occur. 
 
 For $\tilde{g}>1$, only a single event is enough to spread the phase significantly. Decoherence
 coefficients are dominated by noise histories with very low number of switching events. 
 For $\mathcal{D}^{\mathrm{(e)}}_{\lambda}$, those are histories without any switwhing event
 whereas for $\mathcal{D}^{(o)}_{\lambda}$, histories with a single switching event dominate.
 In this strong dephasing limit $\tilde{g} \gg 1$, these expressions can be simplified
 \begin{eqnarray}
 \label{eq:decoherence:telegraphic-even:strong}
 \mathcal{D}^{\mathrm{(e)}}_{\lambda}(t) & = & e^{-\gamma_{\varphi }t/2}\times e^{-i\lambda t}\\
 \label{eq:decoherence:telegraphic-odd:strong}
 \mathcal{D}^{\mathrm{(o)}}_{\lambda}(t) & = & \frac{\gamma_{\varphi}}{2\lambda}\,
 e^{-\gamma_{\varphi}t/2}\,\sin{(\lambda t)}\,.
 \end{eqnarray}
 In the $\tilde{g} \gg 1$ limit, only the "even" contribution survives.
 
 For $\tilde{g}\ll 1$, a large number of switching events is necessary to spread the phase
 $\lambda \int_{0}^tX(\tau)\,d\tau$ significantly. The dephasing time is then much longer
 than the typical waiting time. In this regime, we have:
 \begin{equation}
 \mathcal{D}^{(\mathrm{e}/\mathrm{o})}_{\lambda}(t)\simeq 
 \frac{1}{2}\,e^{-\frac{\lambda^2}{\gamma_{\varphi}}\,t}
 \end{equation}
 leading to a decoherence time of the order of $\gamma_{\varphi}/\lambda^2\gg 
 \gamma_{\varphi}^{-1}$. Both coefficients are equal since having an even
 or an odd number of switchings makes no difference for $t\gg \gamma_{\varphi}^{-1}$.
 
 \medskip
 
 In the present context, the coupling constant $\lambda$ depends on $p_{\pm}$ and $m_{\pm}$. 
 
 For $m_{+}\neq m_{-}$, we focus on the 
 decoherence of a pair of generalized Gea-Banacloche states. In this case, for all values of $p_{\pm}$
 close to $\bar{n}$, the dimensionless coupling is of 
 the order of $\tilde{g}\simeq 2g\sqrt{\bar{n}}/\gamma_{\varphi}\gg 1$
meaning that we are in the $\tilde{g}\gg 1$ regime. In this case, formulas  
 \eqref{eq:decoherence:telegraphic-even:strong} and  \eqref{eq:decoherence:telegraphic-odd:strong}
are directly relevant and the decoherence rate is given by $\gamma_{\varphi}/2$.  

\medskip

 For  $m_{+}=m_{-}=m$, we focus on the decoherence of a single generalized Gea-Banacloche state.
 In this case, for $|p_{+}-p_{-}|\lesssim \sqrt{\bar{n}}$, the dimensionless coupling is
 then of the order of
 $\tilde{g} \simeq mg(p_{+}-p_{-})/\gamma_{\varphi}\sqrt{\bar{n}}$. Its smallest
 non zero value is reached for $|p_{+}-p_{-}|=1$: $mg/\gamma_{\varphi}\sqrt{\bar{n}}$. Therefore, for
 $\bar{n}\ge \bar{n}_{c}$ where $\bar{n}_{c}=(2g/\gamma_{\varphi})^2\gg 1$, the 
 $\tilde{g}\ll 1$ limit can be reached for small enough $|p_{+}-p_{-}|$.  On the other hand, the highest value is 
 $\tilde{g}\simeq mg/\gamma_{\varphi}\gg 1$. 

As long as $\bar{n}\ll \bar{n}_{c}$, which is the case of practical interest here, 
all decoherence coefficients $\mathcal{D}_{\lambda}(t)$ associated with $|p_{+}-p_{-}|
\lesssim \sqrt{\bar{n}}$
are in the $\tilde{g}\gg 1$ regime. But for $\bar{n}\gg \bar{n}_{c}$, this is no
longer the case. In this case, starting with a qubit initially in state $|X_{\pm}\rangle$,
we find that the reduced density operator for the electromagnetic mode is of the form:
\begin{equation}
\label{eq:dephasing:strong-coupling:field-evolution}
\rho_{\mathrm{cav}}(t)=
e^{-\bar{n}}\sum_{p_{+},p_{-}}\frac{\bar{n}^{(p_{+}+p_{-})/2}}{\sqrt{p_{+}!\,p_{-}!}}
e^{-\Gamma_{p_{+},p_{-}}t}\,|p_{+}\rangle\langle p_{-}|
\end{equation}
where $\Gamma_{p_{+},p_{-}}$ is a rate bounded from above by $\gamma_{\varphi}$
and for sufficiently low $p_{+}-p_{-}$, given by:
\begin{equation}
\label{eq:dephasing:strong-coupling:Fock-decoherence-rate}
\Gamma_{p_{+},p_{-}}\simeq 
\frac{\gamma_{\varphi}}{64}\,
\left(\frac{\bar{n}_{c}}{\bar{n}}\right)\,
(p_{+}-p_{p})^2\,.
\end{equation}
The physical interpretation of these results is clear: the condition $\bar{n}\gtrsim \bar{n}_{c}$ means that the 
average number of dephasing jumps within the revival time $4\pi\sqrt{\bar{n}}/g$ is much larger
than one. Since each of these jumps is equivalent to an echo pulse, the phase 
of the electromagnetic mode has a diffusive motion around zero thus leading to decoherence.
In the end, the qubit, incoherent over times scales $\gtrsim \gamma_{\varphi}^{-1}$ 
due to its coupling with its own environment, tends to 
select specific states of the electromagnetic mode. These turn out to be Fock states. The fact that the coherence
between adjacent Fock states decays over a much longer time scale than $\gamma_{\varphi}^{-1}$
(see \eqref{eq:dephasing:strong-coupling:Fock-decoherence-rate})  reflects the time needed for 
a small incoherent object (the qubit) to decohere the mesoscopic coherent field.
  
 \medskip
 
 To summarize, in the regime 
 $g\ll \gamma_{\varphi}$ and $1\ll \bar{n}\ll \bar{n}_{c}$, 
 the main contribution to the atoms + cavity reduced density operator comes
 from quantum trajectories without any quantum jump (decoherence 
 coefficients $\mathcal{D}^{\mathrm{(e)}}_{\lambda}(t)$). 
 
 \subsubsection{Qubit relaxation}
\label{appendix:dynamics:details:relaxation}
 
 Being primarily interested in the Rabi oscillation signals, we shall
 focus on the average value $\langle \sigma^z(t)\rangle$. Eq. 
 \eqref{eq:qubit-observables:population} 
 relates it to the coherences $|X_{+}^{(p)}\rangle \langle X_{-}^{(p)}|$ 
 and $|X_{+}^{(p)}\rangle \langle X_{-}^{(p)}|$.
 Therefore it is sufficient to compute the decoherence coefficient
 $\mathcal{D}_{+-}^{(p,p)}(t)$. Because of the precise form of relaxation 
 quantum jumps, it receives contributions from all four coefficients
 $\mathcal{D}_{\epsilon,\epsilon'}^{(p,p)}(0)$. As a first guess, we might forget about the slow
 phases and compute approximate decoherence coefficients $D_{+-}^{(p,p)}$. In this
 case, the contribution of each pair of path is a well defined pure phase divided by $2^{R[\frak{T}]}$
 where $R[\frak{T}]$ is the total number of relaxation jumps of the underlying quantum trajectory.
 
 \medskip
 
\begin{figure}
 \begin{picture}(0,0)%
\epsfig{file=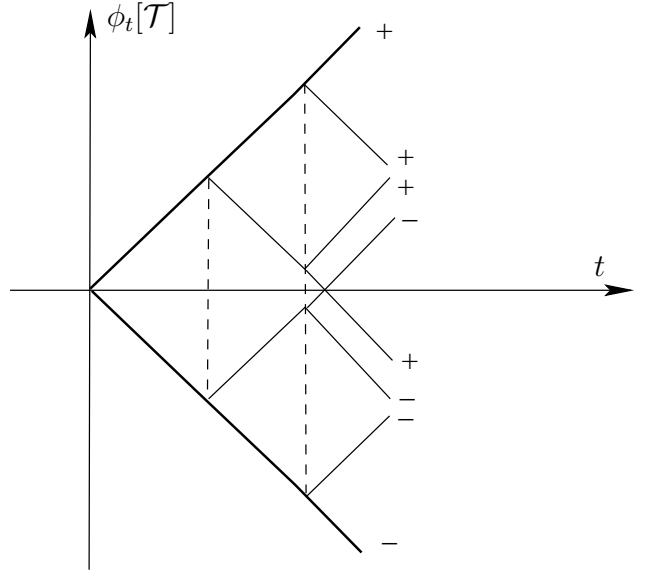}%
\end{picture}%
\setlength{\unitlength}{4144sp}%
\begingroup\makeatletter\ifx\SetFigFont\undefined%
\gdef\SetFigFont#1#2#3#4#5{%
  \reset@font\fontsize{#1}{#2pt}%
  \fontfamily{#3}\fontseries{#4}\fontshape{#5}%
  \selectfont}%
\fi\endgroup%
\begin{picture}(3759,3423)(3101,-4191)
\put(3691,-924){\makebox(0,0)[lb]{\smash{{\SetFigFont{12}{14.4}{\familydefault}{\mddefault}{\updefault}{\color[rgb]{0,0,0}$\phi_t[\mathcal{T}]$}%
}}}}
\put(6605,-2417){\makebox(0,0)[lb]{\smash{{\SetFigFont{12}{14.4}{\familydefault}{\mddefault}{\updefault}{\color[rgb]{0,0,0}$t$}%
}}}}
\end{picture}%
 \caption{\label{fig:relaxation:1} Plots of all pairs of trajectories involved in the evolution of 
 the $|X_{-}\rangle \langle X_{+}|$ 
 coherence having exactly two relaxation jumps. Relaxation jumps are shown as bifurcations.
 The $+$ and $-$ labels indicate which trajectories are relative to the forward and backward Keldysh contours.
 The extremal trajectories that dominate $\langle \sigma^z(t)\rangle$ are indicated with thick lines.}
 \end{figure}

 Consider now a specific pair of these paths $[\mathcal{T}_{+},\mathcal{T}_{-}]$.
 Since we are dealing with $p_{+}=p_{-}$, 
 the phase accumulated between quantum jumps (duration $\Delta\tau$)
 is equal either to $e^{\pm ig\sqrt{p+1} \Delta\tau}$ when the two paths $\mathcal{T}_{+}$
 and $\mathcal{T}_{-}$ are not parallel
 or to $1$ when they are parallel. Thus, one may conveniently associate with all
$[\mathcal{T}_{+},\mathcal{T}_{-}]$
 connecting to $|X_{-}^{(p)}\rangle \langle X_{+}^{(p)}|$ at time
 $t$ a 3-valued noise history: $X(\tau)\in \lbrace 0,\pm ig\sqrt{p+1}\rbrace$
so that the pure phase associated with the pair of paths the is $\exp{(i\int_{0}^tX(\tau)\,d\tau)}$. 
 Starting from the coherence $|X_{+}^{(p)}\rangle
 \langle X_{-}^{(p)}|$, there is one single pair of paths which connects to the same
 coherence and whose associated phase does not
 depend at all on the dates of quantum jumps occuring between $t=0$ and $t$. It
 consists into the pair of extremal paths depicted on fig. \ref{fig:relaxation:1} and corresponds to
$X(\tau)=g\sqrt{p+1}$. Any other pairs of paths will lead to a noise $X(\tau)$
 presenting at least one blip, {\it i.e.} a time interval during 
 which $X(\tau)\neq g\sqrt{p+1}$.
 Any other pair of paths starting from
 $|X_{+}^{(p)}\rangle \langle X_{-}^{(p)}|$ or any pair of paths starting from the diagonal terms 
 $|X_{\epsilon}^{(p)}\rangle\langle X_{\epsilon}^{(p)}|$ ($\epsilon=\pm$) and ending
 on $|X_{+}^{(p)}\rangle\langle X_{-}^{(p)}|$ will exhibit an associated
 noise $\tau\mapsto X(\tau)$ that is not constant.
 
Given a certain noise history $\tau\mapsto X(\tau)$, we shall now sum over 
 all pairs of trajectories  $[\mathcal{T}_{+},\mathcal{T}_{-}]$ associated with a given
 function $\tau\mapsto X(\tau)$. One of them has a minimal number of relaxation jumps
 precisely occuring at the dates where $X(\tau)$ jumps. But 
 of all histories having more relaxation jumps occuring between these specific dates must
 also be taken into account (this is an important difference with the case of pure dephasing studied in the previous 
 paragraph). As shown in appendix \ref{appendix:non-hermitian}, the relaxation jumps 
 stochastic process can be approximated by a renewal process with exponential 
 waiting time distribution: $\psi_{R}(\tau)=\gamma\,e^{-\gamma\tau}$
 where $\gamma=\gamma_{1}/2$.
 Taking into account the factor $1/2$ associated with each relaxation jumps, the resulting
 integration measure for intermediates times $(t_{1},\ldots ,t_{p})$ where $0< t_{1}< \ldots <
 t_{p}< t$ is given by 
 $d\mu(t_{1},\ldots ,t_{p})=e^{-\gamma t/2}\,(\gamma/2)^p\,dt_{1}\ldots dt_{p}$.
 Note that the exponential factor $e^{-\gamma t}$ present in the waiting time distribution is partially
 compensated by the summation of all products of $1/2$ factors associated with relaxation jumps 
 occuring at times different from the $t_{j}$s. The $1/2$ factors associated with jumps
 occuring at times $t_{j}$ are taken into account through $(\gamma/2)^p$.
 
 \medskip
 
 This counting argument makes it clear that the contribution of paths associated with a non
 constant noise history will, comparatively to the case $X(\tau)=X(t)$, involve integrals of phases of
 the form $e^{ig\tau\sqrt{p+1}}$ or $e^{2ig\tau\sqrt{p+1}}$ over $\tau$
 with measure $(\gamma/2) d\tau$, thus leading to a factor proportional to 
 $\gamma/g\sqrt{p+1}$ which, in the mesoscopic regime, is or the order of
 $\gamma / g\sqrt{\bar{n}}$. Besides this, we are in the strong coupling regime and 
 therefore  $\gamma /g\sqrt{\bar{n}}$ is always much smaller than one. Thus, these
 contributions vanish in the regime considered in this paper. 
 
 \medskip
 
 Finally, our analysis shows that the dominant contribution 
to $\mathcal{D}_{+-}^{(p,p)}(t)$ in \eqref{eq:decoherence:double-path-integral}
comes from trajectories $\mathcal{T}_{\pm}$ 
 for which the Fresnel angle $\phi_{\mathcal{T}_{\pm}}$ is monotonous in time. Of course, our 
 argumentation ignored the slow phases that appear in \eqref{eq:decoherence:slow-phases} 
 but it can be checked by explicit computation that 
 our conclusion remains valid for pairs of paths with one or two 
 jumps.
 
Retaining only the contribution of the pair of extremal paths in 
\eqref{eq:decoherence:double-path-integral} and taking slow phases into account, the argumentation of 
 \cite{Degio22} can then be straightforwardly adapted to resum their contribution 
 along these trajectories. This immediatly leads to \eqref{eq:relaxation:decoherence-coefficient} 
 where the $1/2$ prefactor
 in front of $e^{i(\Delta\phi)(\tau)}$ comes from the $1/2$ factor associated with each relaxation jump.
 
 \medskip
 
 As a final comment, the above derivation also makes it clear that depending on the physical quantity
 we are interested in, other pairs of trajectories will contribute. As an example, computing 
 the probability for having the qubit in the $|+\rangle$ or the $|-\rangle$ involves
 coefficients $\mathcal{D}^{(p,p)}_{++}$ $\mathcal{D}^{(p,p)}_{--}$ which are
 dominated by the sum over pair of paths such that $\mathcal{T}_{+}=\mathcal{T}_{-}$.
 
 \subsubsection{Summing over all dissipative processes}
 \label{appendix:dynamics:details:total}

In the presence of dephasing, the extremal trajectories are still the dominant contribution to  
$\langle \sigma^z(t)\rangle$
and the effect of dephasing is an extra $e^{-\gamma_{\varphi}t/2}$ factor. Note that these corresponds to
quantum histories without any dephasing quantum jump. These are expected to provide
the dominant contribution in the low $\gamma_{\varphi}$ limit. 

\medskip

It follows from the previous analysis that the main contribution to $\langle \sigma^z(t)\rangle$
comes from pairs of extremal paths. Since photon losses are independant processes from relaxation jumps,
their contribution can be computed using the formalism developped in Ref. \cite{Degio22}. Obviously, it factors
in front of the relaxation contribution. For $\kappa t\ll 1$, this enables to treat the regime where
photon losses leading to decoherence of the generalized Gea-Banacloche states occur at a 
much higher rate than relaxation and pure dephasing $\kappa \bar{n}\gg \gamma_{1,\varphi}$.


 \end{document}